\documentclass[twocolumn,aps,superscriptaddress,preprintnumbers,floatfix]{revtex4-2}
\usepackage[utf8]{inputenc}
\usepackage[colorlinks=true, allcolors=blue]{hyperref}
\usepackage{graphicx} 
\usepackage{amsmath}
\usepackage{amsfonts}
\usepackage[justification=raggedright,singlelinecheck=false]{caption} 
\usepackage[labelformat=simple]{subcaption}

\DeclareMathOperator*{\argmin}{arg\,min}

\newcommand{\avg}[1]{\left\langle #1 \right\rangle}

\newcommand{\kjrnote}[1]{\textcolor{blue}{#1}}

\newcommand{\hatl}{\widehat{\mathcal{L}}}
\newcommand{\expected}[1]{\E[#1]}
\DeclareMathOperator{\E}{\mathbb{E}}
\newcommand{\ignore}[1]{}

\newcommand{\ot}[1]{\mathcal{O}(\Delta t^{#1})}

\newcommand{\jw}{\text{fluctuation current }}
\newcommand{\jf}{\text{correlation current }}

\begin{document}
\title{
Learning
Stochastic Thermodynamics
Directly from\\
Correlation and Trajectory-Fluctuation Currents}
\author{Jinghao Lyu}
\email{jolyu@ucdavis.edu}
\affiliation{Complexity Sciences Center and Physics and Astronomy Department,
University of California at Davis, One Shields Avenue, Davis, CA 95616}

\author{Kyle J. Ray}
\email{kjray@ucdavis.edu}
\affiliation{Complexity Sciences Center and Physics and Astronomy Department,
University of California at Davis, One Shields Avenue, Davis, CA 95616}

\author{James P. Crutchfield}
\email{chaos@ucdavis.edu}
\affiliation{Complexity Sciences Center and Physics and Astronomy Department,
University of California at Davis, One Shields Avenue, Davis, CA 95616}

\begin{abstract}
Markedly increased computational power and data acquisition have led to growing interest in data-driven inverse dynamics problems. These seek to answer a fundamental question: What can we learn from time series measurements of a complex dynamical system? For small systems interacting with external environments, the effective dynamics are inherently stochastic, making it crucial to properly manage noise in data. Here, we explore this for systems obeying Langevin dynamics and, using currents, we construct a learning framework for stochastic modeling. Currents have recently gained increased attention for their role in bounding entropy production (EP) from thermodynamic uncertainty relations (TURs). We introduce a fundamental relationship between the cumulant currents there and standard machine-learning loss functions. Using this, we derive loss functions for several key thermodynamic functions directly from the system dynamics without the (common) intermediate step of deriving a TUR. These loss functions reproduce results derived both from TURs and other methods. More significantly, they open a path to discover new loss functions for previously inaccessible quantities. Notably, this includes access to per-trajectory entropy production, even if the observed system is driven far from its steady-state. We also consider higher order estimation. Our method is straightforward and unifies dynamic inference with recent approaches to entropy production estimation. Taken altogether, this reveals a deep connection between diffusion models in machine learning and entropy production estimation in stochastic thermodynamics.
\end{abstract}
\date{\today}
\maketitle
\section{Introduction}

Open systems, by their nature, interact with external environments and so experience fluctuations. In small systems, the scale of these fluctuations are often appreciably large when compared to a system's deterministically-controlled dynamics and so, it is imperative to study these fluctuations in nanoscale systems. A straightforward way to investigate small systems is to integrate out these interactions---called fast degree dynamics---which leads to noisy (stochastic) forces. The most famous example is Langevin dynamics, originally used to describe a mesoscopic particle's behavior at a finite temperature \cite{lemons1997paul}. Langevin dynamics has also been used to describe a wide range of stochastic systems from colloid particles and molecules to living cells \cite{bruckner2019stochastic, ariga2018nonequilibrium, fang2019nonequilibrium} and even financial markets \cite{kanazawa2018kinetic}.

Over the last several decades, Langevin dynamics regained attention in the study of systems far from equilibrium, forming a new approach to mesoscopic nonequilibrium physics called \emph{stochastic thermodynamics} \cite{seifert2012stochastic}. This approach elevates standard thermodynamic-average quantities---such as, work, heat, entropy, and the like---to stochastic quantities with values associated with each individual microscopic realization \cite{seifert2005entropy}. The statistics of these quantities can then be computed, recovering traditional thermodynamic averages, but newly enhanced with information about higher moments. The most well-known results are the \emph{fluctuation theorems}, which express an unexpected lawfullness and symmetry in fluctuations \cite{jarzynski1997nonequilibrium, crooks1999entropy}. 

\subsection{Thermodynamic Uncertainty Relations}

\begin{figure}[!t]
    \centering
    \includegraphics[width=\linewidth]{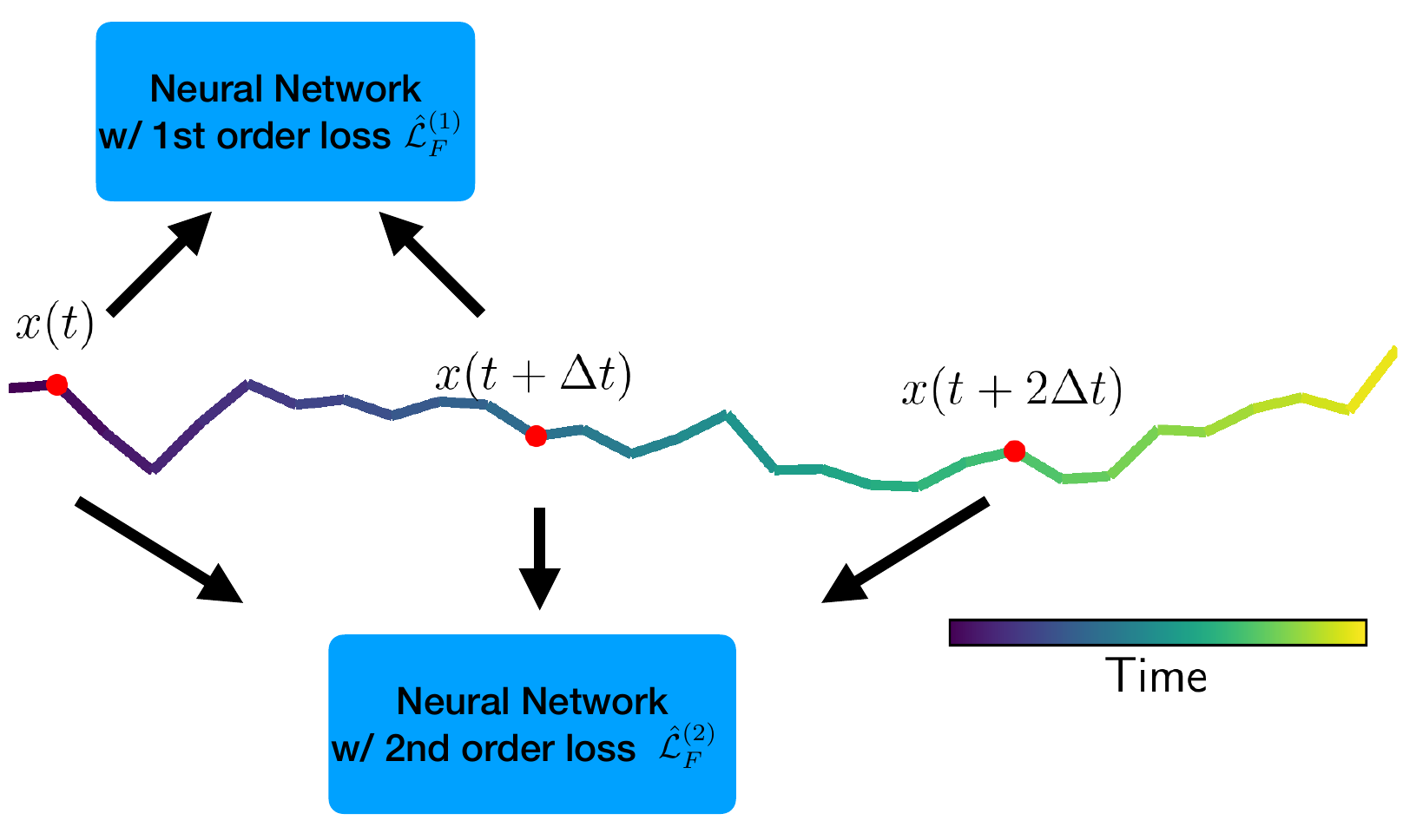}
    \caption{Learning framework: Input data are discrete-time positions: $\{x(t), x(t+\Delta t), x(t+2\Delta t),\dots\}$ and, for underdamped systems, the estimated velocity $\widehat{v}(t) = [x(t+\Delta t) - x(t)]/\Delta t$ is also required. To infer local thermodynamic function $F$ at time $t$---such as force, local entropy production, and diffusion fields---we construct first- and second-order loss functions for training neural networks. First-order estimation relies on two consecutive data points, while second-order estimation uses three. Representative first-order loss functions are summarized in Table~\ref{tab:tablesummary}.}
    \label{fig:cover}
\end{figure}

More recent results in stochastic thermodynamics include the \emph{Thermodynamic Uncertainty Relations} (TURs). The original TUR stated an inequality between the statistics of observable currents and entropy production---between the sum of heat flow and system entropy change under Markovian dynamics in a \emph{non-equilibrium steady state} (NESS) \cite{barato2015thermodynamic}. Since then, TURs were generalized to numerous settings, including time-dependent driven overdamped systems \cite{koyuk2018generalization,koyuk2020thermodynamic,dieball2023direct}, quantum systems \cite{hasegawa2020quantum, hasegawa2021thermodynamic}, systems with feedback control \cite{potts2019thermodynamic}, and underdamped dynamics \cite{van2019uncertainty, dieball2024thermodynamic}.

TUR inequalities can be saturated for certain systems in the short-duration regime. Due to this, they can be used to estimate entropy production, opening up avenues for data-driven exploration of nonequilibrium physics. Entropy production, typically challenging to estimate, can be accessed via current statistics, which are only available from observed trajectory data. Helpfully, TURs build a nontrivial bridge between trajectories and current statistics \cite{busiello2019hyperaccurate,horowitz2020thermodynamic,manikandan2020inferring,van2020entropy, otsubo2020estimating, kim2020learning, kim2022estimating,kwon2024alpha,lyu2024learning}. In parallel with entropy estimation, there are also methods that learn parameters of the Langevin dynamic itself \cite{sarfati2017maximum,perez2018high, frishman2020learning,bruckner2020inferring, bae2024inferring}.

Within the class of overdamped Langevin systems, the mean of the Stratonovich currents appearing in the TUR framework is closely connected to the local entropy production function. Short-time TURs then reduce to a Cauchy–Schwarz inequality \cite{otsubo2020estimating, dieball2023direct}. This equivalence implies that TUR estimation of the local entropy production is mathematically analogous to a learning process that employs cosine similarity as a loss function. 

While TUR methods offer powerful tools for estimating entropy production, they also prompt several intriguing questions. First, it is natural to ask whether the currents can be extended beyond Stratonovich products to quantify thermodynamic quantities other than entropy production. Second, cosine-similarity loss functions focus on vector alignment and may not be suited to regression tasks aimed at estimating scalar quantities. This raises another important question: Is it possible to replace TUR methods with alternatives more intrinsically aligned with estimating local thermodynamic functions, such as employing the \emph{mean squared error} (MSE) as a loss function?

\begin{table*}
    \centering
    \includegraphics[width=\linewidth]{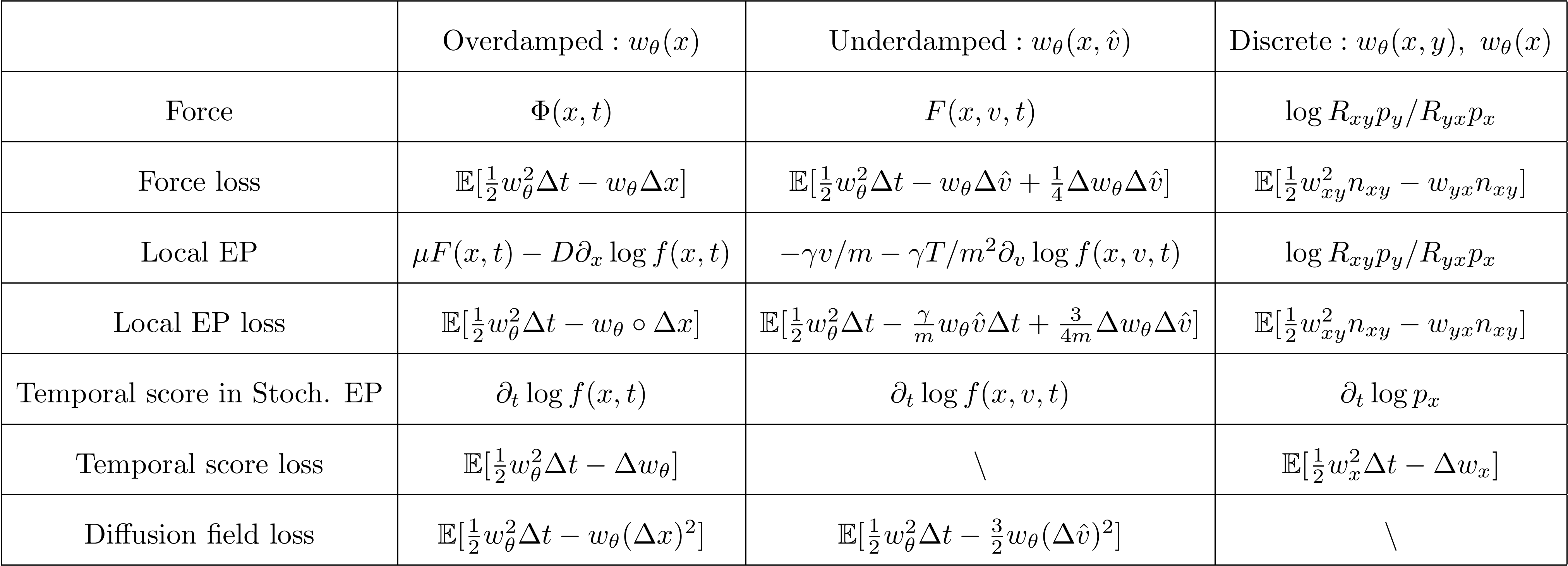}
\caption{MSE-based first-order loss functions for various stochastic systems: a 
    summary of the required data types and the corresponding loss functions for estimating force, local entropy production, temporal score function, and diffusion field in (i) overdamped systems (requiring position trajectories ${x(t)}$), (ii) underdamped systems (requiring position trajectories and estimated velocities ${x(t), \hat{v}(t)}$; see App.~\ref{app:underdamped}), and (iii) discrete-state systems (requiring transition counts ${n_{xy}}$; see App.~\ref{app:Markovian}). Temporal score functions are required to calculate stochastic entropy production. In the underdamped regime, when only position data is available, one cannot estimate stochastic entropy production and so we leave the temporal score loss entry blank; see App.~\ref{app:ud_stochastic_ep}).
    }
\label{tab:tablesummary}
\end{table*}

Despite their promise, TUR methods also have inherent limitations. First, they are sensitive to the choice of time-scale. TURs are saturable only when applied to an infinitesimally-short-time process. To obtain accurate results, the measurement time separation should be near the time scale of the system's dynamics, which is typically unknown in advance. Due to this, it can be complicated to apply TUR methods directly to real world data. Second, TUR methods assume that the stochastic entropy production can be written as an antisymmetric current. Moreover, this is only true when the system is restricted to NESS. Generically, these methods do not have access to trajectory-level stochastic entropy production. Due to this, TURs can only estimate a trajectory's contribution to the \emph{average} entropy production; thereby missing information on higher moments of entropy production. 

\subsection{An Alternative}

The following introduces a learning framework that addresses these challenges by unifying thermodynamic function estimation with machine learning, demonstrating that determining average and stochastic entropy production, diffusion fields, and forces can be reformulated as simple optimizations that minimize a corresponding MSE. We derive loss functions directly from the system dynamics, all the while not requiring the existence and saturability of a TUR. In the case where the measurement-time separation does not match the dynamic time scale, we go on to show that expanding MSEs increases accuracy; we call these expanded MSEs \emph{higher-order loss functions} (See Fig. \ref{fig:cover}).

In numerical experiments, we apply our methods to the nonequilibrium dynamics of a bead-spring model, using a neural network $w_{\theta}$ as universal approximator to estimate thermodynamic functions. This estimates entropy production at the level of individual trajectories for systems driven far from the NESS regime, illustrating that our method still holds for mismatched time-scale scenarios. Accessing entropy production information on a per-trajectory level opens the door to novel kinds of monitoring and measurement device that can detect and potentially correct for previously-invisible large fluctuations when driving between steady states.

More broadly, by deriving loss functions directly from system dynamics, the framework is highly versatile and can be extended to a wide range of thermodynamic functions and systems of interest. For example, we demonstrate that the same underlying principle and procedure can be used to construct loss functions for estimating both force and entropy production and that they apply to underdamped systems, overdamped systems, and Markovian jump processes alike.

Notably, the approach remains effective even in the partially-observed underdamped setting, where velocity information is not directly available and must be inferred. This highlights the method's flexibility and generality. By the same token, we expect that it can be used to find loss functions for other quantities of interest and other dynamical regimes as well. Table~\ref{tab:tablesummary} summarizes the MSE-based loss functions for estimating force (drift) and local entropy production across several common stochastic systems.


\section{Currents as loss functions}\label{overdamped}

Before continuing, several notes about our assumptions and notation are in order. We explicitly consider only systems governed by the overdamped Langevin equation:
\begin{align}\label{eqn:od_master}
    d\boldsymbol{x} &= \boldsymbol{\Phi}(x,t) dt  + \boldsymbol{\sigma} \cdot d\boldsymbol{W}_{t}~,
\end{align}
where $\boldsymbol{\Phi}(x,t)$ is the drift, $d\boldsymbol{W}_{t}$ is an infinitesimal Weiner process, and $\sigma$ is the covariance matrix of the thermal noise. The corresponding Fokker-Planck equation is \cite{pavliotis2014stochastic}:
\begin{align}
    \partial_{t} f(x,t) +\nabla \cdot [f(x,t)\boldsymbol{u}(x,t)] = 0~,
\end{align}
where $f(x,t)$ is the probability distribution and $\boldsymbol{u}(x,t) = \boldsymbol{\Phi}(x,t) - \frac{1}{2} \boldsymbol{\sigma^2} \cdot \nabla \log f(x,t)$ is the probability velocity (or local entropy production rate).
\ignore{and 
\begin{align}
    dx &= v dt  \nonumber\\
    dv &= \Phi(x,v,t) dt+ \boldsymbol{\sigma(x)} dW_{t}
\end{align}
for underdamped dynamics. In some cases, we decompose the underdamped drift $\Phi(x,v,t)$ into damping force $-\gamma v $ and the external force $F(x,v,t)$, i.e., $\Phi=-\gamma v + F$.}
Here, we assume that the covariance $\sigma$ is constant and the diffusion matrix is $\boldsymbol{D}=\frac{1}{2}\boldsymbol{\sigma}^2$.

While we focus on the overdamped case, we also address the underdamped regime as well as Markovian jump processes. These are found in the Appendices \ref{app:underdamped} and \ref{app:Markovian}.

While our method generalizes to higher dimensions without difficulty, we focus on one-dimensional models in the derivation for simplicity.

Finally, our notation is as follows: $\expected{\cdot}$ refers to the theoretical expected value and $\avg{\cdot}$ to the average value from data; the loss function $\mathcal{L}$ without accent refers to the theoretical value from an infinite amount of data and the loss function $\widehat{\mathcal{L}}$ refers to the empirical value calculated from a data set $\mathcal{D}$.

\subsection{Lesson from machine learning}

In machine learning the unified way to construct a loss function is very simple. Suppose the data set is $\mathcal{D}=\{z_{i}\}_{i=1}^{N}$, distributed according to $f(z)$. When we use a model $w_{\theta}(z)$ with parameters $\theta$ to learn a function $F(z)$, a common approach is to use a variation of stochastic gradient descent to minimize the following $L^2$ loss function, the MSE \cite{hastie01statisticallearning}:
\begin{align}\label{eq:L2_norm}
    \mathcal{L}_{F}^{\text{MSE}}(\theta):= \E {[w_{\theta}(z)-F(z)]^2}
    ~.
\end{align}
In practice, one approximates this expectation by averaging over the data set $\mathcal{D}$:
\begin{align}
    \widehat{\mathcal{L}}_{F}^{\text{MSE}}(\theta,\mathcal{D})&= \avg{[w_{\theta}(z)-F(z)]^2}\nonumber \\&
    =\frac{1}{N}\sum_{z_{i}\in \mathcal{D}}{[w_{\theta}(z_{i})-F(z_{i})]^2}~.
\end{align}
If the ground truth $\{F(z_{i})\}_{z_{i}\in \mathcal{D}}$ is known, then the problem becomes one of fine tuning model architecture, optimization parameters, and the like. However, when inferring an unknown $F(z)$, the problem is more challenging as one cannot compute $\widehat{\mathcal{L}}_{F}^{\text{MSE}}(\theta,\mathcal{D})$ exactly using the data $\mathcal{D}$ and the model $w_{\theta}(z)$ alone. Instead, its must be approximated. Expand $\widehat{\mathcal{L}}_{F}^{\text{MSE}}(\theta,\mathcal{D})$ into three qualitatively different parts:
\begin{align}\label{app_eqn:ml_difference_expansion}
    \hatl_F^{\text{MSE}}({\theta, \mathcal{D}}) &= \avg{w_{\theta}(z)^2} 
    - \avg{2 w_{\theta}(z) F(z)} + \avg{F(z)^2}
    ~.
\end{align}
The last term can always be neglected as it is $\theta-$independent. And so, parameter optimization only affects the first two terms in Eq. \eqref{app_eqn:ml_difference_expansion}. This suggests always using the equivalent loss function:
\begin{align}\label{eq:L2_loss}
    \mathcal{L}_{F}(\theta):= \E {\left[\frac{1}{2} w^2_{\theta}(z)\right]- \E\left[w_\theta(z)F(z)\right]}
    ~.
\end{align}
The first term depends only on the model. And so, it can always be calculated easily. However, the second term requires special attention. Though one \emph{cannot} know $\{F(z_{i})\}_{z_{i}\in \mathcal{D}}$ in advance, all one needs is a quantity that \emph{can be} approximated from the data set $\mathcal{D}$ whose average is \emph{proportional to} $w_{\theta}(z) F(z)$. Estimating this correlative term is the core of our approach.

Let's specialize to time series data $z=(x,t)$. If we assume access to a continuous time series and group  data per trajectory, the loss function takes the form of:
\begin{align}\label{eq:L2_loss_time_series}
    \mathcal{L}_{F}(\theta)= \E {\left[\int dt\frac{1}{2} w^2_{\theta}(x,t)\right]- \E\left[\int dtw_\theta(x,t)F(x,t)\right]}
    .
\end{align}
For real data, with discrete-time trajectories, the integrals become finite sums. If we have $N$ discrete-time trajectories, each of length $n$ steps, the dataset is given by $\mathcal{D}=\{x_{i}(a \Delta t)\}_{a=0,i=1}^{n,N}$, where $x_{i}(a \Delta t)$ represents the position of the $i$-th trajectory at time step $a$. Assume the goal is to learn a time-dependent overdamped drift $F(x,t)=\Phi (x,t)$. Eq. \eqref{eqn:od_master} tells us that:
\begin{align}
    \expected{w_{\theta}(x,t) \Phi(x,t)\Delta t} = \expected{w_{\theta}(x,t) \Delta x } + \mathcal{O}(\Delta t^2)~,
\end{align}
where $\Delta x = x(t+\Delta t)-x(t)$ is the change in position. For finite data with discrete time steps, we then expect:
\begin{align}
    \avg{w_{\theta}(x,t) \Phi(x,t)}\Delta t \sim \avg{w_{\theta}(x,t) \Delta x } + \mathcal{O}(\Delta t^2)
    ~.
\end{align}

Substituting this into Eq. \eqref{eq:L2_loss_time_series} leads to the following approximate loss function that can be calculated from the model $w(x)$ and the data $\mathcal{D}$ alone:
\begin{align}\label{eqn:od_loss_force}
    \hatl^{(1)}_\Phi = \avg{\sum_{a=0}^{n-1}\left[\frac{1}{2}w_{\theta}(x,a\Delta t)^2\Delta t -  w_{\theta}(x,a\Delta t)\Delta x(a\Delta t)\right]} ~,
\end{align} 
where $\Delta x(a\Delta t)=x((a+1)\Delta t)-x(a\Delta t)$ for each trajectory at time $t$. The superscript ${(1)}$ means the loss function agrees with the true $\mathcal{L}$ up to $\mathcal{O}(\Delta t)$.

This approach has several advantages. First, it simplifies the task of finding the loss function to a simple question: To estimate $F$, find a functional that has the same average as $w F$ (up to a multiplicative constant) that can be estimated from the data $\mathcal{D}$. Next, it introduces the ability to evaluate and extend the inference accuracy. That is, for agreement beyond leading order in $\Delta t$, the method is clear: estimate the term in question to a higher order. Finally, it allows theoretical connections to previous TUR methods for inference, which we discuss next.

\subsection{Unification with TUR estimates}

We can treat these loss functions as averages of cumulant currents. Starting here, for clarity we suppress the model's explicit $\theta$ dependence. For a general function, $F(x,t)$, the theoretical value of the loss function in Eq. \eqref{eq:L2_loss_time_series} for long trajectory data is:
\begin{align}
    \mathcal{L}_F &=\int dt \int dx[ \frac{1}{2} w(x,t)^2  -  w(x,t)F(x,t) ] f(x,t) \\&=\expected{J_{w}} - \expected{J_{F}}~,
\end{align}
where $J_{w} =1/2 \int_{\Gamma} w(x,t)^2 dt$ and $J_{F} = \int_{\Gamma} w(x,t)F(x,t) dt$ are two integrals along the trajectory and $f(x,t)$ is the distribution. The function $w$ that minimizes the $L^2$ norm is the same as that yielding the minimal expected value difference between the $J_{w}$ and $J_{F}$ currents. We refer to these as the \emph{\jw} and the \emph{\jf}, as they capture trajectory fluctuations in the weight function $w$ and its correlation with the target function $F$, respectively. 

From this perspective, we view the saturating $w$ as the result of a variational principle. That is, by inspection, one sees that the variation with respect to $w$ is stationary only when $w(x,t)=F(x,t)$ with the extreme value being $-\frac{1}{2}\E[F^2]$. While the appearance of this variational principle is intriguing on its own, perhaps hinting at a physical intuition behind the method, let's now compare this approach to TUR-based estimation. 

 The primary key to the flexibility of the approach is an expanded notion of what a ``current" can be, defining it as the product of some arbitrary local weight function with any other local thermodynamic function. In contrast, TURs use a more targeted current definition. Broadly stated, the TUR defines currents through the Stratnovich product $w(x,t) \circ dx$. General TURs, which bound entropy production through the Cr\'amer-Rao inequality, provide valuable theoretical insights but are typically not saturable. However, in the short time limit, these TURs can often be saturated and thus used for estimation \cite{van2020entropy}. Due to the targeted choice of the Stratonovich product current, its ensemble average is proportional to $\E[u w]$. In short, the current defined in the TUR can be seen as a particular example of a \jf: $J_F$ for $F=u$. Since the average entropy production is related to $\E[u^2]$, the TUR can be reduced to a statement of the Cauchy-Shwarz inequality between $u$ and $w$ \cite{dieball2023direct, lee2023multidimensional, lyu2024learning}.

In our language, this means employing cosine similarity:
\begin{align}
\mathcal{L}_{u}^{\text{cosine}}=\frac{ \expected{w^2}\expected{u^2} }{ \expected{wu}^2}
\end{align}
as the loss function. Cosine similarity saturates under proportionality rather than equality, when $w(x,t)=c u(x,t)$. However, the constant can be removed after saturation by calculating the ratio $\expected{u^2}/\expected{wu}^2$. Thus, our $L^2$ method yields the same result as a TUR estimation without going through the intermediate steps of deriving or choosing a TUR and considering the tightness of the inequality.

One can imagine, in light of this, a ``TUR-based" method to estimate the drift $\Phi$. Our considerations above show that to leading order in the timestep, a current defined as $ w\Delta x$ yields an ensemble average of $\Delta t \E[w\Phi]$. And, this suggests a Cauchy-Shwarz Inequality for short times along the lines of:
\begin{align}
\E[\Phi^2]\E[w^2]\geq \E[w\Phi]^2\sim \E[w\Delta x]^2/\Delta t^2
~.
\end{align}
As above, saturation occurs with proportionality between $w$ and $\Phi$, leading to an analogous cosine similarity loss function for estimating the drift. And again, the proportionality can be removed post saturation. 

There is, however, a subtle difference between $L^2$ and TUR estimation that is worth pointing out. Given a weight function $w$ and ground truth $F$, we have the following inequalities:
\begin{align}
\expected{F^2} \geq 2\expected{F w}-\expected{w^2}
\end{align}
for the $L^2$ norm and:
\begin{align}
\expected{F^2} \geq \frac{\expected{wF}^2}{\expected{w^2}}
\end{align}
for the cosine similarity. A third inequality:
\begin{align}
\frac{\expected{wF}^2} {\expected{w^2}} \geq 2\expected{F w}-\expected{w^2}
\end{align}
suggests that the cosine similarity used in the TUR provides a better lower bound on $\expected{F^2}$ compared to the $L^2$ norm. While this is true given the same $w$, it is not a fair and complete comparison when determining which loss function is superior. In particular, it does not account for how the training process under these two loss functions affects the closeness of the weight $w$ to the ground truth $F$. With infinite data, the weight functions $w$ using these two losses converge to the same underlying function $F$, assuming the optimization is successful and the model has sufficient capacity. Our primary interest is, then, how quickly and reliably one can train $w$ to convergence. Likely, this is better achieved using $L^2$ norm-based loss functions, if we look at the overwhelming majority of machine-learning loss functions for evidence. $L^2$ loss is largely preferred over cosine similarity for regression tasks both in the literature and in practice.

No matter the chosen loss, the estimation problem is reduced to the following principle: the extent to which we can learn the local function $F(x,t)$ from data is the extent to which we can approximate the average \jf $\expected{J_F}$ from that data. The extent to which we can do this depends on the dynamics at play and the data. While this method works for long-time trajectories when the model $w$ is expressive enough to capture time dependence, we focus on short-time trajectories and currents to study local function estimation at a given time. To distinguish the short single-time currents we use lower case letters:
\begin{align}
J_F(t+dt)-J_F(t) = j_F dt=w(x,t)F(x,t)dt
    ~.
\end{align}


\subsection{Loss functions in overdamped dynamics}

The last section derived the loss function using the It\^o current for the drift $\Phi$ as an illustrative example. The following derives loss functions for other local quantities in stochastic thermodynamics. For simplicity of notation, we show derivations for overdamped dynamics only. We also derived loss functions for underdamped systems and Markovian jump processes; the relevant derivations and loss functions are in Appendices \ref{app:underdamped} and \ref{app:Markovian}.

The average entropy production, the primary target of TUR-based estimation, monitors how irreversible a process is. The average entropy production rate at time $t$ is proportional to the expected value of $u(x,t)^2$. To learn the probability velocity $u(x,t)$, we want to build a loss function $\mathcal{L}_u(w)$ that satisfies $u = \argmin_w \mathcal{L}_u(w)$ at time $t$. From the previous section, we expect the loss function with infinite data to be:
\begin{align}
\mathcal{L}_u &=dt \int dx [\frac{1}{2} w(x)^2 - w(x)u(x, t)] f(x,t)\nonumber\\ 
&= \expected{j_w dt} - \expected{j_{u}dt}~.
\end{align}

The problem to solve is straightforward: Estimate the average $u$-\jf from data alone. The infinitesimal current expectation value is:
\begin{align}
\expected{j_{u}dt}&= dt
\int dxf(x,t)w(x)[\Phi(x,t)\nonumber
    -\frac{1}{2}\frac{\partial_{x}f(x,t)}{f(x,t)}\sigma^2 ]\\
    &=\int dx f(x,t)[w(x,t)\Phi(x,t) dt \nonumber +\frac{1}{2}\partial_{x}w(x) \sigma^2dt]\\
    &=\expected{w(x) \cdot dx}+ \expected{\frac{1}{2} \partial_{x}w(x) \sigma^2 dt} \nonumber\\
    & =\expected{ w(x) \circ dx} ~,
\end{align}
where we integrate by parts from the first line to the second and convert between It\^{o} and Stratonovich integrals in the last line. It is worth noting that one of the relevant terms takes the same form as the learning score function used in a reverse diffusion model \cite{hyvarinen2005estimation, hyvarinen2007some, song2021scorebased}. Moreover, we see the appearance of a Stratonovich current-based loss function we mentioned earlier, which (along with several variants) has been commonly used to estimate the entropy production in previous TUR studies \cite{van2020entropy,otsubo2020estimating,frishman2020learning,kim2020learning,otsubo2022estimating}. 

With data $\mathcal{D}$, the corresponding $L^2-$based discretized loss function from $t$ to $t+\Delta t$ is, then:
\begin{align}\label{app_eqn:od_ent_infer_loss}
    \hat{\mathcal{L}}_{u}^{(1)} = \avg{\frac{1}{2}w(x)^2 \Delta t - \frac{1}{2}[w(x)+w(x+\Delta x)] \Delta x}
    ~,
\end{align}
where $\Delta x=x({t+\Delta t})-x(t)$. Suppose the minimum of $\hatl_{u}^{(1)}$ is achieved when $w(x) =u^{(1)}(x)$, then the average estimated entropy rate is:
\begin{align}
\widehat{\Sigma}=\frac{1}{D}\avg{{u^{(1)}(x)}^2}
~.
\end{align}
With infinite data and an infinitesimal time step $dt$, this loss function yields $u^{(1)}(x) = u(x)$. With a finite amount of discrete-time data, however, we need to keep in mind that the loss function is only correct up to $\mathcal{O}(\Delta t)$. We will discuss the difference between $u^{(1)}(x)$ and $u(x)$ shortly when addressing higher-order inference.

\ignore{, i.e.,
\begin{align}
    \mathcal{L}_{u}^{(1)}(w,\mathcal{D}) &= \avg{\frac{1}{2}w(x)^2 \Delta t - \frac{1}{2}[w(x+\Delta x)+w(x)] \Delta x} \nonumber\\
    &= \avg{\frac{1}{2}w(x)^2  - \frac{1}{2}w(x) u(x,t)} \Delta t+ \mathcal{O}(\Delta t^2) 
\end{align}
We can also find the contribution from the $\mathcal{O}(\Delta t^2)$ \kjrnote{I have not worked though this yet. Can we do it?}. Then: 
\begin{align}
     &\avg{C_{w}^{(1)}(t_{i})} = \avg{\frac{1}{2}w(x_{i})^2 \Delta t - \frac{1}{2}[w(x_{i+1})+w(x_{i})] \Delta x_{i} }\\
     &=\langle\frac{1}{2}w(x_{i},t_{i})^2 \Delta t - w(x_{i},t_{i}) \Delta x_{i} -\frac{1}{2} \partial_{x}w(x_{i},t_{i}) \Delta x_{i}^2 \nonumber\\
     &-\frac{1}{4} \partial_{xx}w(x_{i},t_{i}) \Delta x_{i}^3-\frac{1}{12} \partial_{xxx}w(x_{i},t_{i}) \Delta x_{i}^4\rangle + \mathcal{O}(\Delta x^5)~.
\end{align}
To find the $\mathcal{O}(\Delta t^2)$, we have
\begin{align}
    \avg{\Delta x_{i}} &\sim \Phi(x_{i},t_{i}) \Delta t + \frac{1}{2} \Phi(x_{i},t_{i}) \partial_{x}\Phi(x_{i},t_{i}) \Delta t^2 \\
    \avg{\Delta x_{i}^2} &\sim \sigma^2 \Delta t +  \Phi(x_{i},t_{i}) \Delta t^2 + \sigma^2 \partial_{x}\Phi(x_{i},t_{i}) \Delta t^2\nonumber \\
    &+\frac{1}{4}\sigma^2 \partial_{xx}\Phi(x_{i}) \Delta t^2\\
    \avg{\Delta x_{i}^3} &\sim 3\sigma^2 \Phi(x_{i},t_{i})  \Delta t^2 \\
    \avg{\Delta x_{i}^4} &\sim \sigma^4 \Delta t^2 
\end{align}
}

Rather than considering average entropy production, we can also estimate the entropy production of each trajectory---the \emph{stochastic entropy production} \cite{seifert2005entropy}:
\begin{align}
\label{eqn:def_od_ent}
    \sigma_{\Gamma} &=\int_{\Gamma} [u(x,t)\circ dx + \partial_{t} \log f (x,t) dt]~.
\end{align}
For a NESS, $u(x,t)$ is sufficient for capturing the stochastic entropy production since the temporal score function $\partial_t\log f$ vanishes. However, in a non-NESS this term must be accounted for. Using TURs only, it is simply not possible to capture the $\partial_{t}\log f dt$ current contribution. Specifically, TURs assume that currents have the form of $\int w(x,t)\circ dx$, where $w(x,t)$ is a weight function. However, we need not make any such assumption. And, with our framework, we have effectively reduced the problem to quadrature: To learn the additional contribution temporal score function, approximate:
\begin{align}
    \expected{j_{d_{t}\log f}dt} &= dt\int dx f(x,t)w(x)\partial_{t}\log f(x,t) \nonumber\\
    &=dt \int dx w(x)\partial_{t}f(x,t) \nonumber\\
    &=dt  \frac{d}{dt}\int dx f(x,t)w(x) \nonumber\\
    &=\expected{w(x)}_{t+dt}-\expected{w(x)}_{t}
    ~,
\end{align}
The corresponding discretized loss function from $t$ to $t+\Delta t$ is:
\begin{align}
    \widehat{\mathcal{L}}^{(1)}_{d_{t}\log f}&= \avg{\tfrac{1}{2}w(x)^2 \Delta t}- \avg{w(x)}_{t+\Delta t} + \avg{w(x)}_{t}\nonumber\\
    &= \avg{\tfrac{1}{2}w(x)^2 \Delta t- w(x+\Delta x) +w(x)}~.
\end{align}
This loss function allows one to learn non-NESS contributions to the entropy production per-trajectory. We can also learn the variance of the thermal noise $\sigma^2$, the loss function is:
\begin{align}
    \mathcal{L}_{\sigma^2}&= dt \int dx [\frac{1}{2} w^2 - w\sigma^2] f(x,t)~.
\end{align}
Here, we assume the diffusion field is not position-dependent. Notice that the infinitesimal displacement $dx$ satisfies $(dx)^2 \sim \sigma^2 dt $. The discretized loss function is:
\begin{align}\label{eqn:loss_diffusion_field_empi}
    \widehat{\mathcal{L}}^{(1)}_{\sigma^2}=\avg{\frac{1}{2}w^2\Delta t-w (\Delta x)^2}~.
\end{align}
from which we can learn $\sigma^2$. If $\sigma^2(x)$ depends on the location, we can generalize the scalar weight $w$ in the loss function in Eq. \eqref{eqn:loss_diffusion_field_empi} to a function $w(x)$.

\subsection{Loss functions at higher order}

The superscript ${(1)}$ highlights that the empirical loss function $\widehat{\mathcal{L}}^{(1)}$ only considers contributions to order $\mathcal{O}(\Delta t)$. If the data is generated with a time step of $dt\sim\Delta t$---matching the measurement interval $\Delta t$---the first-order inference results are sufficiently accurate. However, in real data, the measurement interval $\Delta t$ may not coincide with $dt$. In such cases, expanding the empirical loss functions to higher orders can compensate by increasing the accuracy of inference. Here, we demonstrate the mechanism of higher-order inference. Appendix \ref{app:higher_order} provides a more exhaustive mathematical treatment. 

 Recall that we extract the drift information from the average of $\Delta x$:
 \begin{align}
 \E(\Delta x) = \Phi(x,t)f(x,t)\Delta t +\mathcal{O}(\Delta t^2)
    ~.
 \end{align}
 With $\Delta t$ being sufficiently small, the neural network can accurately capture the drift function $\Phi(x,t)$ with the $\ot{2}$ term vanishing.
 
 If $\Delta t$ is not small enough, we can correct for the discretization error. Taking into account the next order:
 \begin{align}
 \E(\Delta x) = \Phi(x,t)f(x,t)\Delta t +c_{2}(x,t)\Delta t^2 + \mathcal{O}(\Delta t^3)
 ~,
 \end{align}
 where $c_{2}(x,t)$ is a function of $x$ and $t$. To remove the effect from $\mathcal{O}(\Delta t^2)$ term, we introduce $\Delta_{2} x=x(t+2\Delta t)-x(t)$, for which:
 \begin{align}
 \E(\Delta_2 x) = \Phi(x,t)f(x,t)2\Delta t +c_{2}(2\Delta t)^2 + \mathcal{O}(\Delta t^3)
 ~.
 \end{align}
 To cancel out the $\mathcal{O}(\Delta t^2)$ term, we have:
 \begin{align*}
 \E \left[2\Delta x-\tfrac{1}{2}\Delta_{2}x \right]=\Phi(x,t)f(x,t)\Delta t+ \mathcal{O}(\Delta t^3)
 ~,
 \end{align*}
 leaving only contributions that are $\mathcal{O}(\Delta t^3)$ and smaller. This cancellation is identical to the method used in finite difference schemes \cite{fornberg1988generation}. It motivates us to construct the higher-order loss function for the drift $\Phi$:
 \begin{align}
 \label{eqn: loss_function_force_2nd}
     \hatl_{\Phi}^{(2)} = \avg{\frac{1}{2}w(x)^2 \Delta t - w(x)(2\Delta x-\tfrac{1}{2}\Delta_{2}x)} ~,
 \end{align}
 which is correct up to $\ot2$.
 
 We can easily construct even higher-order loss functions, though this comes at the cost of requiring more steps in trajectories. Similarly, other local functions can be estimated with high orders. Here, we give the second-order loss functions for probability velocity $u$, $\partial_{t}\log f$, and thermal bath variance $\sigma^2$:
 \begin{align}
 \label{eqn:loss_function_other_2nd}
     \hatl_{u}^{(2)} & =
     \left\langle
     \tfrac{1}{2} w(x)^2 \Delta t - [w(x+\Delta x)+w(x)] \Delta x \right. \nonumber \\
     & \left. \qquad  +\tfrac{1}{4}[w(x+\Delta_{2} x)+w(x)]\Delta_{2} x
     \right\rangle
     \\
     \hatl_{d_{t}\log f}^{(2)} & = \avg{\tfrac{1}{2}w(x)^2 \Delta t - 2 \Delta w+\tfrac{1}{2} \Delta_{2}w} \\
     \hatl_{\sigma^2}^{(2)} & =\avg{\tfrac{1}{2}w^2\Delta t-w[ 2(\Delta x)^2-\tfrac{1}{2}(\Delta_2 x)^2]}
     ~.
 \end{align}
Derivations and additional details on higher-orders  are found in Appendix \ref{app:higher_order}.
\ignore{
where we assume the drift $\Phi$ is time-independent. If we use the same loss function $\hatl_{\Phi}^{(1)}$, the optimal weight $w^{*}$ will learn $w^{*}(x)=\Phi(x)  +\left(\frac{1}{2}  \Phi(x) \partial_{x}\Phi(x) + \frac{1}{4}\sigma^2\partial_{xx}\Phi(x)   \right)\Delta t$. To solve $\Phi(x)$ perturbatively, we can expand $\Phi(x)$ as a series of $\Delta t$: $\Phi(x)=C_{0}(x)+ C_{1}(x) \Delta t + \dots$ and solve $C_{0}(x)$, $C_{1}(x)$ accordingly. The results are
\begin{align}
    \Phi(x) =& w^*(x) - \frac{1}{2} \Delta t \left( w^*(x)\partial_{x}w^*(x)+\frac{1}{2} \sigma^2 \partial_{xx}w^{*}(x)\right) \nonumber\\
    &+ \mathcal{O}(\Delta t^2)~.
\end{align}
To infer at higher order, we still use loss function $\hatl_{\Phi}^{(1)}$ to train the network and only need to include higher order correction afterward. For score $\partial_{x}\log f$, we can consider a loss function to learn the score. From integral by parts, we have $\E(\frac{1}{2}w^2+\partial_{x}w)=\E(\frac{1}{2}w^2-w\partial_{x}\log f )$. The corresponding empirical function $\avg{\frac{1}{2}w^2+\partial_{x}w}$ does not include the data at the next step. This is similar to the loss function used in diffusion model training. This empirical loss function would learn the score function at all orders, though the derivative term makes it harder to train using back-propagation. We list the details of higher order in Appendix.}

We will not analyze how measurement error affects our estimation scheme in detail, as it falls beyond our main scope. However, accounting for the effects of measurement noise in our framework is straightforward, as its behavior differs from that of thermal noise. We can mitigate the impact of measurement noise by incorporating longer trajectory data points, just as is the higher-order inference scheme. We illustrate with the simple example of learning thermal variance $\sigma^2$. Suppose that the measurement error (denoted $\epsilon$) is time independent and has $0$ mean and standard deviation $\sigma_{\epsilon}$. Then $\expected{(\Delta x)^2}=\sigma^2 \Delta t+ \sigma_{\epsilon}^2+ \ot2 $ and $\expected{(\Delta_{2} x)^2}=2\sigma^2 \Delta t+ \sigma_{\epsilon}^2+\ot2 $. To eliminate the effect of noise, we use the difference between $ \expected{(\Delta x)^2}$ and $\expected{(\Delta_{2} x)^2}$ to learn the thermal variance.

\begin{figure*}
    \centering
    \begin{subfigure}{.48\linewidth}
        \centering
        \caption{}
        \includegraphics[width=\linewidth]{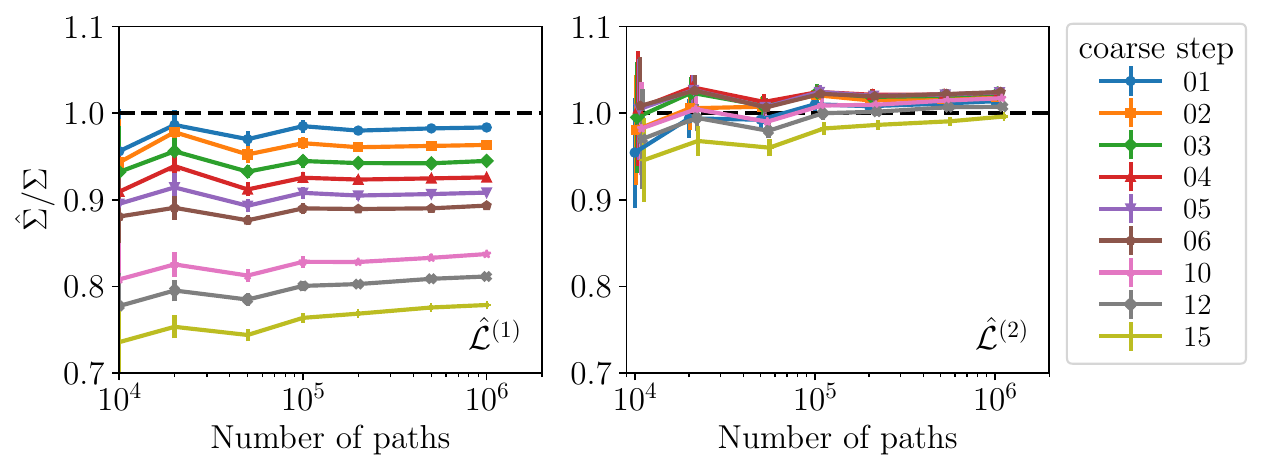}
        \label{fig:ent_N_dis2}
    \end{subfigure}
    \hfill
    \begin{subfigure}{.48\linewidth}
        \centering
        \caption{}\includegraphics[width=\linewidth]{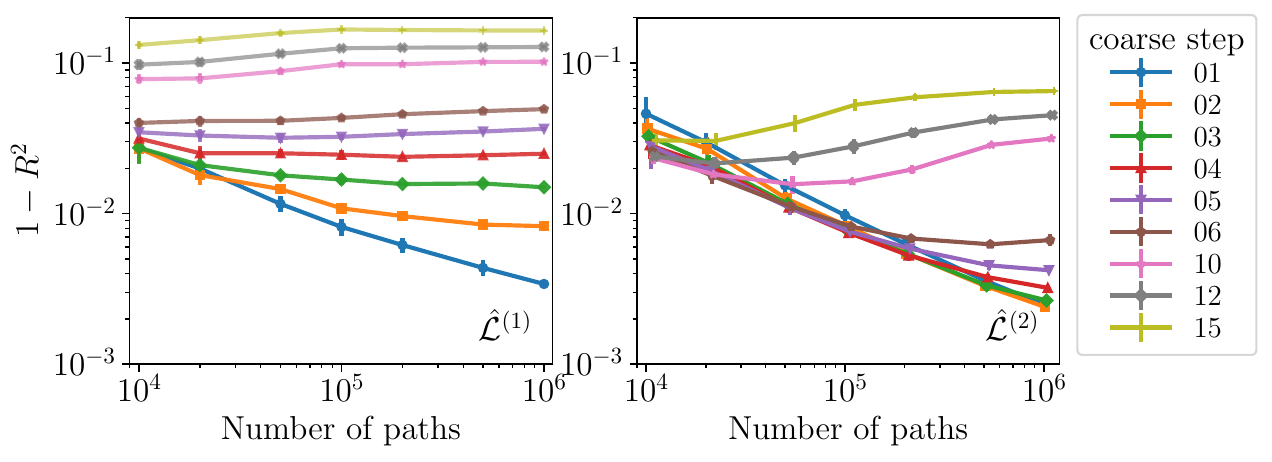}
        \label{fig:r2_N_dis2}
    \end{subfigure}
    \hfill
    \begin{subfigure}{.48\linewidth}
        \centering
        \caption{}
        \includegraphics[width=\linewidth]{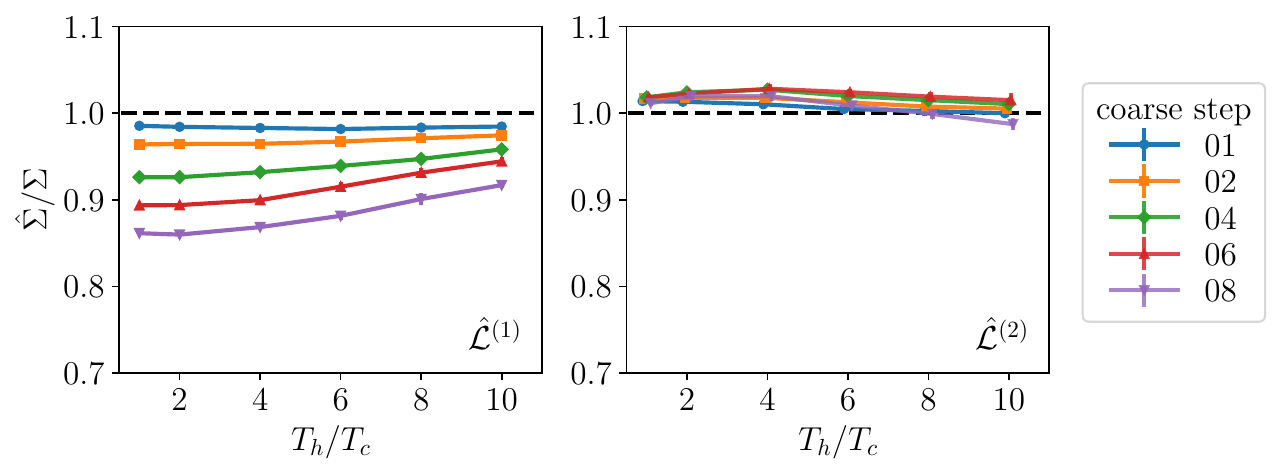}
        \label{fig:ent_temp_dis2}
    \end{subfigure}
    \hfill
    \begin{subfigure}{.48\linewidth}
        \centering
        \caption{}
        \includegraphics[width=\linewidth]{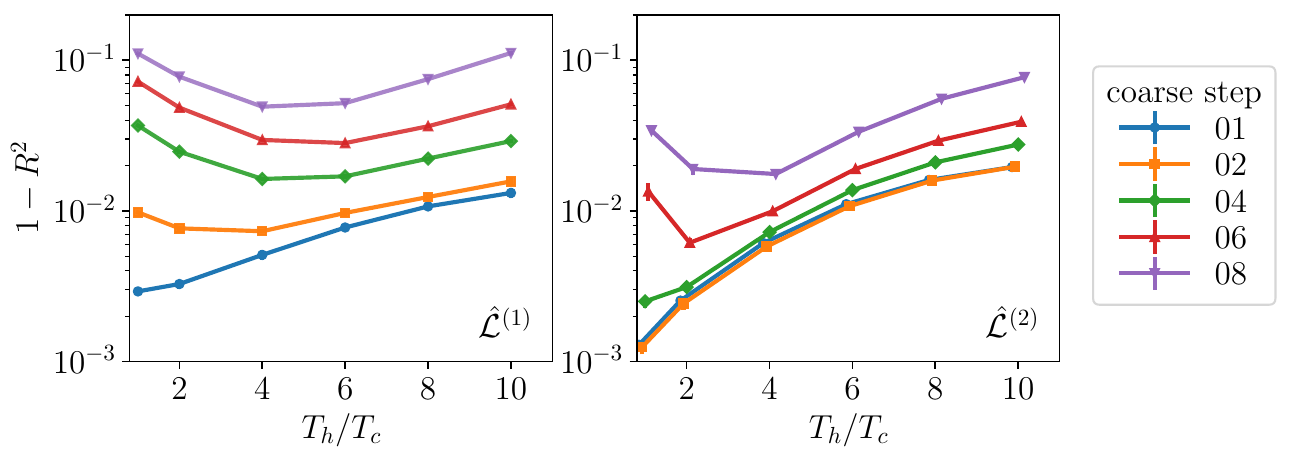}
        \label{fig:r2_temp_dis2}
    \end{subfigure}
    \begin{subfigure}{.48\linewidth}
        \centering
        \caption{}
        \includegraphics[width=\linewidth]{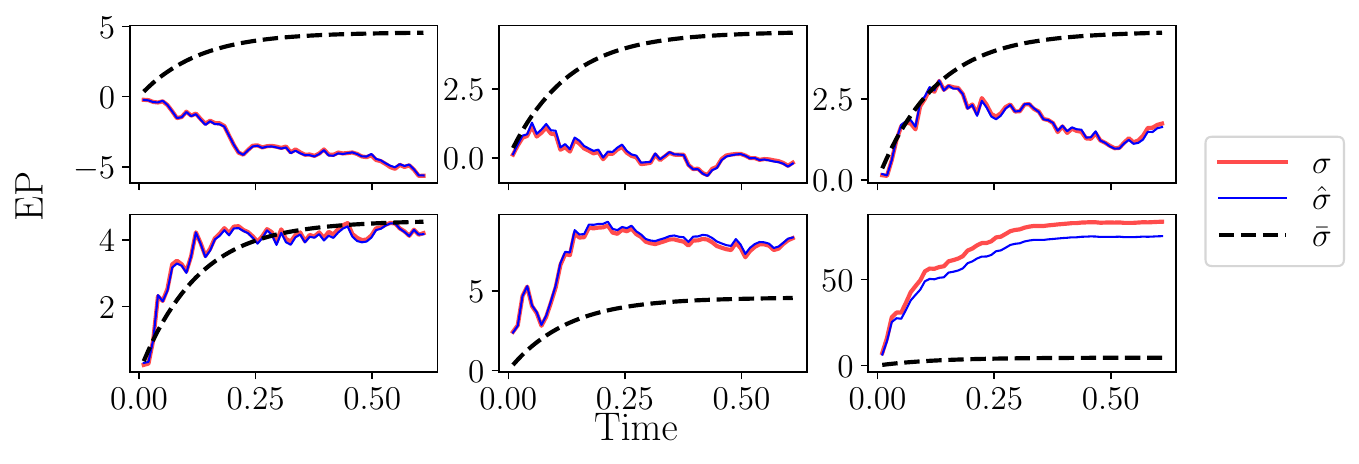}
    \label{fig:nn_stochastic_ep_no_coarse}
    \end{subfigure}
\hfill
    \begin{subfigure}{.48\linewidth}
    \centering
       \caption{}
       \includegraphics[width=\linewidth]{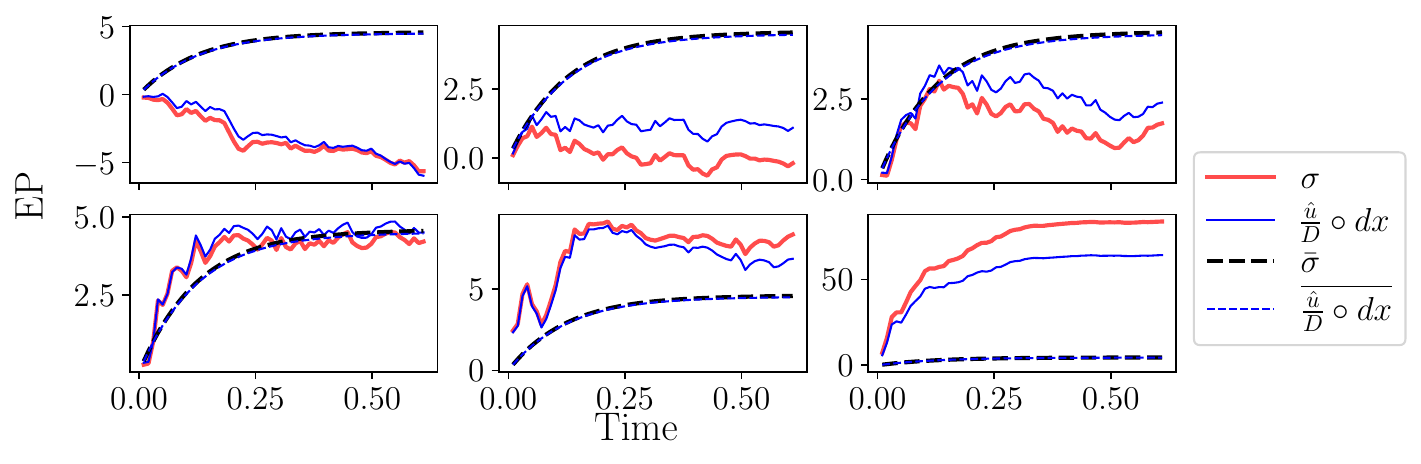}
    \label{fig:nn_u_stochastic_ep_compare_no_coarse}
    \end{subfigure}
\caption{Entropy production inference with five beads-spring model. (a) 
    The estimated relative entropy production $\hat{\Sigma}/\Sigma$ as a function of the number of paths using with different coarse steps. The left/right are the relative entropy results with the first/second order loss functions. (b) The $1-R^2$ score between trajectory's theoretical and estimated stochastic entropy productions as functions of number of paths. The left/right are the results with the first/second order loss functions. (c) The relative entropy production results at different temperature ratios with different types of $10^6$ coarse-grained trajectories. (d) The corresponding $1-R^2$ score between theoretical stochastic entropy production and estimated entropy production of validating trajectories. (e) Entropy production of six chosen trajectories from the bead-spring model at the temperature ratio $T_{h}/T_{c}=2$. The theoretical stochastic entropy productions are computed from the definition: the sum of change in the surprisal and the heat along the trajectory. The dashed line is the average entropy production as a function of time $t$. (f) The role of $\partial_{t}\log f$ in stochastic entropy production. The stochastic entropy production (red) and Stratonovich integral of $u$ (blue) along trajectories are shown. The dashed lines are average entropy production and $\int D^{-1}\cdot  \boldsymbol{u} \circ d\boldsymbol{x}$.
    }
\label{fig:results_dis2}
\end{figure*}

\section{Numerical Results}

Extensive previous efforts investigated overdamped Langevin dynamics including estimating forces, average entropy production, and stochastic entropy production for NESS states \cite{van2020entropy,otsubo2020estimating,frishman2020learning,kim2020learning,otsubo2022estimating}. In contrast, this section focuses on stochastic entropy production outside of the NESS regime---a quantity that is inaccessible using TUR methods. For this estimation, we use the loss functions for $\boldsymbol{u}$ and $\partial_{t}\log f$ derived in the previous section.

The stochastic entropy production $\sigma_\Gamma$ of trajectory $\Gamma$ is given in Eq. 
\eqref{eqn:def_od_ent}. During estimation, one often ignores the term $\partial_{t} \log f (x,t)$ since it does not contribute to the average entropy production for any time evolution. However, it does contribute to higher-order moments of the entropy production. For example, in one-dimensional free diffusion starting with a Gaussian distribution with variance $\sigma^2$, the first two moments of the entropy production from $0$ to $t$ are:
\begin{align}
    \expected{\Sigma}&=\frac{1}{2}\log \frac{2 Dt+\sigma^2}{\sigma^2} ~,\\
    \text{Var}(\Sigma)&= \frac{2Dt}{2Dt +\sigma^2}~,
\end{align}
where $D$ is the diffusion constant. If we ignore the $\partial_{t}\log f$ term, the results are:
\begin{align}
    \expected{\int_{\Gamma}u\circ dx}&=\frac{1}{2}\log \frac{2 Dt+\sigma^2}{\sigma^2}~,\\
    \text{Var} \left(\int_{\Gamma}u\circ dx \right)&= \log \frac{2 Dt+\sigma^2}{\sigma^2}~.
\end{align}
As we can see, $\int_{\Gamma}u\circ dx$ can be used to compute average entropy production, but fails to capture its variance.
\ignore{We first learn $\partial_{t}\log f$ in the free diffusion. 
The theoretical result for a free diffusion starting with a Gaussian distribution with variance $\sigma^2 $ is 
\begin{align}
    \partial_{t}\log f(x,t)=-D\frac{(2Dt - x^2 + \sigma^2)}{(2Dt + \sigma^2)^2}~.
\end{align}
}

Let's explore the new framework for the nonequilibrium thermodynamics of the bead-spring model---a commonly-used benchmark for testing entropy production estimation methods. The model belongs to the class of \emph{linear models} that obey a Langevin equation of the form:
\begin{align}
    d\boldsymbol{x} = A \boldsymbol{x}dt + \sqrt{2D}\cdot d\boldsymbol{B}_{t}
    ~,
\end{align}
where $\boldsymbol{x}$ is an $n$-dimension state vector, $A$ is an $n\times n$ matrix, and $D$ is a positive diagonal matrix. We consider five beads at different temperatures connected in series by setting:
\begin{align}
    A = \mu\begin{pmatrix}
        -2k & k & 0 & 0 &0\\
          k & -2k & k &0 &0\\
          0& k & -2k & k  &0\\
          0& 0& k & -2k & k \\
          0 & 0 & 0 &k &-2k
    \end{pmatrix}
    ~,
\end{align}
where $\mu$ is the mobility of beads and $k$ is the spring stiffness \cite{li2019quantifying}. We set the diffusion constant matrix to $D=\frac{1}{4} \mu k_{\mathrm{B}}T $diag$\{(4T_{c},3T_{c}+T_{h},2T_{c}+2T_{h},T_{c}+3T_{h},4T_{h})\}$ to model contact with five different thermal baths, whose temperatures linearly vary from $T_{c}$ to $T_{h}$. For our simulations, we set all parameters to $1$: $\mu = k =k_{\mathrm{B}} =1$. We generate $10^6$ trajectories with initial conditions sampled from a non-NESS distribution and time step $\Delta t=10^{-2}$. The initial distribution is Gaussian with mean $0$ and covariance matrix $\text{diag}(1,2,3,4,5)$. To estimate stochastic entropy production, we train $2$ Multilayer Perceptrons (MLPs) to learn probability velocity $\boldsymbol{u}$ and $\partial_{t}\log f$, respectively, at each time step. In each training, we allocate $80\% $ of the trajectories to the training set and $20\%$ percent to the validation set. We transfer the MLP parameters from the previous time step to initialize the next, accelerating the training process. More training details are in Appendix \ref{app:trainingdetails}.

We tested the performance of first- and second-order loss functions, with results from this experiment shown in Fig. \ref{fig:results_dis2}. All estimates were computed from the withheld validation trajectory set. We first demonstrate the mean entropy production estimation results allowing for access to different numbers of trajectories and time coarse-grainings. For ``coarse step'' $a$, the neural network can only access the data points $\{\boldsymbol{x}_{ia\Delta t}\}_{i=0}^{[60/a]}$. Figure \ref{fig:ent_N_dis2} shows that the usual first-order loss function produces poor estimation of the average entropy production when the coarse step $a > 10$. Whereas the second-order loss function performs significantly better for all coarse steps shown.

We illustrate trajectory-level estimation in Fig. \ref{fig:r2_N_dis2} by showing $1-R^2$. The second-order estimation (right) produces better estimates of stochastic entropy production. In Fig. \ref{fig:r2_N_dis2}, we observe nonmonotonic behavior with coarse step $a > 10$ using both the first- and the second-order loss functions. We believe this is caused by the neural networks' inability to learn the correct local thermodynamic functions due to the large coarse step. Instead, they are converging to functions that are close to the ground truth because they share trajectory statistics at the coarse grained points. With increasing data, the models better approximate this other ground truth, and so their performance on our ground truth degrades. Regardless, the second-order loss function provides more accurate estimation results than first order, even when applied to high coarse-step data.

Figures \ref{fig:ent_temp_dis2} and \ref{fig:r2_temp_dis2} show testing entropy production estimation at different temperature ratios where $T_{c}$ is set to be 1. Entropy production of six different trajectories are presented in Figs. \ref{fig:nn_stochastic_ep_no_coarse} and \ref{fig:nn_u_stochastic_ep_compare_no_coarse} to show our training results using non-coarse data. The theoretical entropy production is computed from the definition directly: $\sigma_{\Gamma}=\Delta s_{\Gamma} +D^{-1}\cdot \int_{\Gamma} \boldsymbol{F}\circ d\boldsymbol{x}$ and the estimated entropy production are computed from the integral: $\hat{\sigma}_{\Gamma}=\int_{\Gamma}(- \partial_t \log f dt + D^{-1}\boldsymbol{u}\circ d\boldsymbol{x})$. Our neural networks accurately capture the stochastic entropy production of both low and high entropy production paths. Recall that the average of $\partial_{t}\log f$ is zero, allowing us to discard this term when estimating the average entropy production. As shown in Fig. \ref{fig:nn_u_stochastic_ep_compare_no_coarse}, this results in the average of entropy production and $u$ integral coinciding. However, the $\partial_{t} \log f$ is non-negligible at the trajectory level.

Figure \ref{fig:nn_u_stochastic_ep_compare_no_coarse} shows the discrepancy between the stochastic entropy production and the $u$ integral contribution. Appendix \ref{app:results_diffdis} presents the results with different initial distributions.

\section{Discussion and conclusion}

Data-driven learning problems are now attracting significant attention not only in classical systems but quantum systems as well \cite{huang2020predicting,gebhart2023learning}. These promise to greatly expand our understanding of far-from-equilibrium thermodynamics processes.

Here, we introduced a unified framework for learning local functions in stochastic thermodynamics, applicable even to partially observable systems. We employed simple MSE as loss functions, approximating these functions using only available trajectory data. Indeed, to infer the underlying thermodynamic functions in a setting with unobservable degrees of freedom, one must estimate loss functions from available data. We demonstrated the primary importance in this of good estimates of \jf: a functional of an observed trajectory that links a candidate model's behavior to that of the observable of interest. We also showed how to expand loss functions appropriately in terms of the measurement separation $\Delta t$ to account for scenarios where the dynamical time scale does not match the measurement time scale. We demonstrated that this ensures estimating more accurate representations of the stochastic system. As a proof of concept we illustrated our method by learning trajectory-level entropy production for a bead-spring model driven far out of its steady state. Beyond serving as an example use case for the new estimation method, that investigation highlighted the role that $\partial_{t}\log f$ plays in non-NESS stochastic entropy production.

We briefly place our results in the context of other recent efforts. The most straightforward approach is to use kernel methods. \cite{lander2012noninvasive, li2019quantifying} learn the probability distribution through kernel methods and estimate the entropy production. These direct methods also require detailed information about the system’s dynamics and suffer from the curse of dimensionality, making them inefficient or impractical for high-dimensional systems. Regarding to force inference, \cite{frishman2020learning, bruckner2020inferring} use kernel method to learn Langevin dynamics. The method require the prior selection of orthogonal basis functions. Another large family of entropy production estimation methods relies on TURs, which use ratios of statistical currents as loss functions. TURs do not need dynamic information. However, TUR methods can only probe average entropy production. They are blind to non-NESS contributions to the stochastic entropy production. One can define entropy production via the Kullback–Leibler (KL) divergence and estimate the forward and backward trajectory probability \cite{roldan2010estimating, roldan2012entropy, kim2020learning, kim2022estimating}. At short time duration, expanding the KL divergence leads to the same loss functions as those we derived here. Compared to relying on KL divergence, our method using MSE has smooth well-behaved derivative behaviors and can be easily generalized to long time trajectories. Moreover, our method extends beyond functions that must be written in terms of a KL divergence. A new method---the \emph{variance sum rule}---was  proposed recently \cite{di2024variancesum, di2024variance} that estimates NESS average entropy production from the definition $\expected{\boldsymbol{F}\circ d\boldsymbol{x}}$, expressing it in terms of variance of force and positions. Unfortunately, it only applies to systems in a NESS.

In this work, we also showed how to extend our inference scheme to higher orders. Previous entropy production estimations require well-resolved data ($\Delta t \sim dt$) to avoid additional terms proportional to $\Delta t$. We showed that these entropy production estimation schemes break down around $\Delta t\sim 10dt$. Our inference method eliminates additional contributions from higher-order terms in $\Delta t$, resulting in a more accurate representation of the dynamics. A similar strategy has also been used in diffusion models to improve the accuracy and stability of the reverse diffusion process and to generate high-quality samples \cite{lu2022dpm}.

Let's close by suggesting several future directions that are now possible using the new estimation method. First and foremost, it is important to test our method using non-steady state experimental data. A good candidate is the optically-trapped colloidal particle \cite{di2024variance}. In the higher-order inference scheme, the results are less sensitive to the measurement time separation $\Delta t$. Our method promises to provide more accurate probes of system irreversibility than currently available and will serve as a basis for detecting fluctuations in entropy production. Second, it is of interest to explore and test loss functions for other stochastic systems. Active Brownian systems come immediately to mind \cite{boffi2024deep, huang2025entropy}. Another interesting direction would be to develop analogous methods for open thermodynamic quantum systems. While multiple TURs have been formulated for open quantum systems \cite{hasegawa2021thermodynamic, van2022thermodynamics, van2023thermodynamic, gong2022bounds, landi2024current} adapting our method will provide new probes for detecting quantum irreversibility. 

\section*{Acknowledgments}

The authors thank the Telluride Science Research Center for its hospitality during visits and the participants of the Information Engines workshop there for their valuable feedback. This material is based on work supported by, or in part by, the U.S. Army Research Laboratory and U.S. Army Research Office under Grant No. W911NF-21-1-0048.

\bibliography{ref}

\clearpage

\appendix
\section{Neural Net Training} \label{app:trainingdetails}

We use Fully Connected Neural Networks (FCNNs) with ReLU activation functions to learn thermodynamic local functions. They are trained using the Adam optimizer with the following hyper-parameters: learning rate $10^{-4}$, weight decay $10^{-5}$, and batch size $4096$. To avoid overfitting, we employ early stopping. We train FCNNs using the training set and monitor the loss functions on the validation set at each training epoch. If the loss computed on the validation set does not decrease to a new minimum within $5$ consecutive epochs, we terminate training. During training, we load the FCNN parameters from time step $i$ to the FCNN parameters from time step $i+1$ to speed training. We train our FCNNs on an Apple M1 MacStudio using the pytorch `mps' backend.

\section{Overdamped Entropy Production}

The details of our overdamped Brownian particle with mobility $\mu$ in a constant diffusion field---\emph{homogeneous diffusion}---are as follows. The corresponding Langevin dynamic is given by a stochastic differential equation (SDE):
\begin{align}\label{eqn:od_SDE}
    d\boldsymbol{x} = \boldsymbol{\Phi}( \boldsymbol{x},t) dt  + \boldsymbol{\sigma}\cdot d\boldsymbol{W}_{t}
    ~,
\end{align}
where $\frac{1}{2}\boldsymbol{\sigma}^2  =\boldsymbol{D}$---the diffusion matrix. The particle distribution $f(x,t)$ evolution is governed by the Fokker-Planck equation:
\begin{align}
    \partial_{t} f(\boldsymbol{x},t)+\nabla\cdot[\boldsymbol{u}(\boldsymbol{x},t)f(\boldsymbol{x},t)]=0
    ~,
\end{align}
where $\boldsymbol{u}(\boldsymbol{x},t)=\boldsymbol{\Phi}(\boldsymbol{x},t)-\boldsymbol{D}\cdot\nabla\log f(\boldsymbol{x},t)$ is the probability current.
We first describe the trajectory-dependent physical quantities in overdamped Langevin dynamics. For a single trajectory $\{\boldsymbol{x}_t\}_{t=0}^{t=\tau}$, the heat flux into the thermal environment divided by temperature is:
\begin{align}
    q_{\Gamma} = \boldsymbol{D}^{-1}\cdot\int_{\Gamma} \boldsymbol{\Phi}(\boldsymbol{x},t) \circ d\boldsymbol{x}
    ~,
\end{align}
and the surprisal change is:
\begin{align}
    \Delta s_{\Gamma}&=-\log f(\boldsymbol{x}_{\tau},\tau) + \log f(\boldsymbol{x}_{0},0) \nonumber\\
    &=-\int_{\Gamma}[\nabla\log f(\boldsymbol{x},t)\circ d\boldsymbol{x} -\partial_{t}\log f(\boldsymbol{x},t) dt]
    ~.
\end{align}
The entropy production for a single trajectory is:
\begin{align}
    \sigma_{\Gamma} &=  q_{\Gamma} + \Delta s_{\Gamma}\nonumber \\
    &=\int_{\Gamma} \left\{ [\boldsymbol{D}^{-1}\cdot{\boldsymbol{\Phi}(\boldsymbol{x},t)}
    -\nabla\log f(\boldsymbol{x},t)]\circ d\boldsymbol{x} \nonumber \right. \\
    & \left. \qquad -\partial_{t}\log f(\boldsymbol{x},t) dt \right\}
    ~.
\end{align}
If the system is in its NESS, $\partial_{t} \log f(\boldsymbol{x},t)$ vanishes and the trajectory-based entropy production is:
\begin{align}
    \sigma_{\Gamma} &=\boldsymbol{D}^{-1}\cdot\int_{\Gamma} [\boldsymbol{\Phi}(\boldsymbol{x},t)-\boldsymbol{D}\cdot\nabla\log f(\boldsymbol{x},t)]\circ d\boldsymbol{x}\nonumber\\
    &=\boldsymbol{D}^{-1}\cdot\int_{\Gamma} \boldsymbol{u}(\boldsymbol{x},t)\circ d\boldsymbol{x}
    ~.
\end{align}

\subsection{Discretized Langevin equation and currents } \label{app:higher_order}

To learn system dynamics from data, we must align the discrete-time version of the SDE with the discretized data. The formal solution of the Langevin SDE in one dimension is:
\begin{align}
    x_{\Delta t} = x_{0} + \int_{0}^{\Delta t} \Phi(x_s,s) ds + \sigma W_{\Delta t}
    ~,
\end{align}
which has the form of Dyson series. We can obtain the solution to this equation by iteration:
\begin{align}
     x_{\Delta t} & = x_{0} + \int_{0}^{\Delta t} \Phi \left (x_{0}+\int_{0}^{s}\Phi(x_{s'},s')ds'+\sigma W_{\Delta t},s \right) ds \nonumber\\
     & \qquad + \sigma W_{\Delta t}
     ~.
\end{align}
Up to order $\ot{1}$:
\begin{align}\label{app_eqn:1st_od_eqn}
   x_{\Delta t} &= x_{0} +  \Phi(x_0,0) \Delta t + \sigma W_{\Delta t} + \mathcal{O}(\Delta t^{3/2})
   ~,
\end{align}
where the $\sigma W_{\Delta t}$ is $\ot{1/2}$. The discretized version of an Ito current with weight $w(x)$ is:
\begin{align}
    J_{I}(w) &= w(x) \Delta x\nonumber \\
    & = w(x)(\Phi(x,t) \Delta t + \sigma W_{\Delta t})~ + \mathcal{O}(\Delta t^{3/2})
\end{align}
and $\expected{J_{I}(w)}= \expected{w(x)\Phi(x,t) \Delta t}+\ot{2}$. Here, we use the fact that terms proportional to $\Delta t^{3/2}$ must include $\int W_{t}dt$ or $\int tdW_{t}$ both of which have expected value zero, leaving the contribution starting from $\ot{2}$. The drift function $\Phi(x,t)$ appears in the Ito current expected value. To extract the drift function $\Phi(x,t)$ at time $t$, we build a quadratic loss function:
\begin{align}\label{app_eqn:od_loss_current}
    \mathcal{L}_{\Phi}^{(1)}&=\E[{ \frac{1}{2}w(x)^2 \Delta t-w(x) \Delta x}]\nonumber \\
    & = \E\left[\frac{1}{2}w(x)^2 \Delta t-w(x)[\Phi(x,t) \Delta t + \sigma W_{\Delta t}]\right]\nonumber\\
    & \qquad + \ot{2}\nonumber \\
    &= \E\left[\frac{1}{2}[w(x) -\Phi(x,t)]^2 \right]\Delta t+\expected{\Phi(x,t)^2}\Delta t+\ot{2}
    ~.
\end{align}
The minimum is achieved when:
\begin{align}\label{app_eqn:minimum_force_1st}
    \argmin_{w}\mathcal{L}_{\Phi}^{(1)}=\Phi(x,t)+\ot1~.
\end{align}
In this derivation, we only use the property that the average behavior of $\Delta x$ conditioned on $x$ at time $t$---the expectation of Eq. \eqref{app_eqn:1st_od_eqn}---is:
\begin{align}
   \E[\Delta x|x,t] = \Phi(x,t) \Delta t + \mathcal{O}(\Delta t^2)~.
\end{align}
To compute the order $\ot{1}$ term in Eq. \eqref{app_eqn:minimum_force_1st},
we expand $\Delta x$ beyond $\mathcal{O}(\Delta t)$ using the Ito-Taylor expansion. Up to $\mathcal{O}(\Delta t^2)$:
\begin{align}\label{app_eqn:secondodlagevin}
     x_{\Delta t}-x_{0} &=  \sigma W_{\Delta t} +  \Phi(x_0,0) \Delta t  +\sigma \partial_{x}\Phi(x_{0},0)\int_{0}^{\Delta t}ds  W_{s}\nonumber\\
     & \quad +\frac{1}{2} \Delta t^2 \Phi(x_0,0) \partial_{x}\Phi(x_0,0) + \frac{1}{2} \Delta t^2 \partial_{t}\Phi(x_0,0)\nonumber \\
     & \quad + \tfrac{1}{2}\sigma^2\partial_{xx}\Phi(x_0,0)  \int_{0}^{\Delta t} ds W_{s}^2 + \mathcal{O}(\Delta t^{5/2})~.
\end{align}
The average of $x(\Delta t)$ up to $\mathcal{O}(\Delta t ^2)$ is:
\begin{align}
    \expected{\Delta x|{x,t}} &=   \Phi(x,t) \Delta t  \nonumber\\
     & \quad +\tfrac{1}{2} \Delta t^2 \Phi(x,t) \partial_{x}\Phi(x,t) + \frac{1}{2} \Delta t^2 \partial_{t}\Phi(x,t)\nonumber \\
     & \quad + \frac{1}{4}\Delta t^2\sigma^2\partial_{xx}\Phi(x,t)   + \mathcal{O}(\Delta t^{3})~.
\end{align}
So up to $\ot{1}$:
\begin{align}
    &\argmin_{w}\mathcal{L}_{\Phi}^{(1)}=\Phi(x,t)\nonumber\\
    +&\tfrac{1}{2} \Delta t\Big( \Phi(x,t) \partial_{x}\Phi(x,t) +  \partial_{t}\Phi(x,t)
     + \frac{1}{2}\sigma^2\partial_{xx}\Phi(x,t)\Big)
     \nonumber\\
     +&\mathcal{O}(\Delta t ^2)~.
\end{align}
As expected, the loss function $\mathcal{L}_\Phi^{(1)}$ yields an answer that deviates from the actual $\Phi$ by a term that is at the first order in $\Delta t$. We can include one more step from trajectories to eliminate this additional contribution, similar to how one approximates a derivative with a finite difference. For one step, we have:
\begin{align}
    \expected{\Delta x|x,t}=\Phi(x,t) \Delta t + C_{2}(x,t) \Delta t^2+\ot{3}
    ~,
\end{align}
where $C_{2}(x,t)=\frac{1}{2}\Phi(x,t) \partial_{x}\Phi(x,t) + \frac{1}{2} \partial_{t}\Phi(x,t)+ \frac{1}{4}\sigma^2\partial_{xx}\Phi(x,t) $. If we extend the trajectory one step further:
\begin{align}
    \expected{\Delta_2 x|x,t}=\Phi(x,t) (2\Delta t) + C_{2}(x,t)(2 \Delta t)^2+ \ot{3}
    ~,
\end{align}
where $\Delta_2 x = x(t+2\Delta t )-x(t)$. Using these two conditional probabilities, we eliminate the contribution at $\ot{2}$:
\begin{align}
    \frac{1}{2} \big( 4\expected{\Delta x|x,t}-\expected{\Delta_2 x|x,t} \big)= \Phi(x,t) + \ot{3}
    ~.
\end{align}
This leads to the loss function $\mathcal{L}_{\Phi}^{(2)}$:
\begin{align}
    &\mathcal{L}_{\Phi}^{(2)}= \expected{\frac{1}{2} w(x)^2\Delta t-w(x)(2\Delta x-\frac{1}{2}\Delta_{2}x)}\nonumber \\
    &=\expected{\frac{1}{2} w(x)^2 \Delta t-w(x)\Phi(x,t)\Delta t}+\ot{3}\nonumber \\
    &=  \E\left[\frac{1}{2}[w(x) -\Phi(x,t)]^2 \right]\Delta t+\expected{\Phi(x,t)^2}\Delta t+\ot{3}
    ~.
\end{align}
The minimum is achieved when:
\begin{align}\label{app_eqn:minimum_force_2nd}
    \argmin_{w}\mathcal{L}_{\Phi}^{(1)}=\Phi(x,t)+\ot{2}
    ~.
\end{align}
We can also extend this method to further, continually improving the estimation's accuracy. Table \ref{tab:finite_diff} gives the coefficients for higher-order estimation up to $\ot{4}$.

\begin{table}[h]
    \centering
    \scalebox{0.95}{
    \begin{tabular}{|c|c|c|c|c|c|c|c|c|c|c|}
        \hline
         Accuracy & $x(t)$ & $x(t+\Delta t)$ & $x(t+2\Delta t)$ & $x(t+3\Delta t)$ & $x(t+4\Delta t)$   \\
        \hline
         $\ot{1}$ & $-1$ & $1$ & & &  \\
        \hline
         $\ot{2}$ & $-3/2$ & $2$ & $-1/2$ & & \\
        \hline
         $\ot{3}$ & $-11/6$ & $3$ & $-3/2$ & $1/3$ &   \\
        \hline
         $\ot{4}$ & $-25/12$ & $4$ & $-3$ & $4/3$ & $-1/4$   \\
        \hline
    \end{tabular}
    }
    \caption{Loss function coefficients for the drift $\Phi$ inference.}
    \label{tab:finite_diff}
\end{table}

To infer the probability velocity $u(x,t)$, we also derive higher-order loss functions. Consider a Stratonovich product current with weight $w$---$J_{II}(w) = w \circ \Delta x$:
\begin{align}
    J_{II}(w) 
     & = \tfrac{1}{2}[w(x+\Delta x) +w(x)] \Delta x\nonumber\\
    &=w(x)\sigma W_{\Delta t}+ w(x)\Phi(x) \Delta t \nonumber \\
    & \quad + \tfrac{1}{2} \partial_{x}w(x) \sigma^2 W_{\Delta t}^2  + \ot{3/2}
    ~.
\end{align}
The corresponding expected value is:
\begin{align}
    &\expected{J_{II}(w)} \nonumber \\
    & \quad = \E[w(x)\Phi(x,t) + \frac{1}{2}\partial_{x}w(x)\sigma^2 ] \Delta t + \ot{2}~ \nonumber \\ 
    & \quad =  \E \left[w(x)[\Phi(x,t) - \frac{1}{2}\partial_{x} \log f(x,t)\sigma^2] \right] \Delta t + \ot{2} \nonumber\\
     &\quad =  \E \left[w(x)u(x,t) \right] \Delta t + \ot{2}
     ~,
\end{align}
where we integrate by parts from the first line to the second line and $f(x,t)$ is the probability distribution. To extract the function $u(x,t)$, the quadratic loss function is:
\begin{align}
    \mathcal{L}_{u}^{(1)}= \E \left[ \tfrac{1}{2}w^2(x)\Delta t - \tfrac{1}{2}[w(x+\Delta x) +w(x)] \Delta x \right]
    ~.
\end{align}
Again, this loss function gives the correct answer only up to first order since we only took the expectation of $J_{II}$ up to leading order. Once again, to obtain accuracy at the next order, we must remove the contribution using the position at $t+2\Delta t$:
\begin{align}
    \expected{w(x)\circ \Delta x} &= \expected{w(x)u(x,t)} \Delta t + C^{u}_{2}(x,t) \Delta t^{2} \nonumber\\
    & \quad + \ot{3} \nonumber\\
    \expected{w(x)\circ \Delta_{2} x} &= \expected{w(x)u(x,t)} 2\Delta t + C^{u}_{2}(x,t) 4\Delta t^{2}\nonumber\\
    & \quad + \ot{3}
    ~,
\end{align}
where $w(x)\circ \Delta_{2} x = 1/2[w(x+\Delta_{2}x )+w(x)] \Delta_{2}x$. To cancel out $C_{2}^{u}(x,t)$:
\begin{align}
    2\expected{w(x)\circ \Delta x} & -\frac{1}{2}\expected{w(x)\circ \Delta_{2} x} \nonumber \\
    & = \expected{w(x)u(x,t)} \Delta t + \ot{3}~.
\end{align}
And the corresponding loss function is:
\begin{align}
    \mathcal{L}_{u}^{(2)} = \E[\frac{1}{2} w(x)^2 \Delta t -2w(x)\circ \Delta x+\frac{1}{2}w(x)\circ \Delta_{2} x]~.
\end{align}
While not necessary for inference, we can also calculate the coefficient $C_{2}^{u}(x,t)$ using a series expansion: 
\begin{align}
     & \mathcal{L}_{u}^{(1)} = \avg{\frac{1}{2}w(x)^2 \Delta t - \frac{1}{2}[w(x)+w(x+\Delta x)] \Delta x }\\
     &=\left\langle\frac{1}{2}w(x)^2 \Delta t - w(x) \Delta x -\frac{1}{2} \partial_{x}w(x) \Delta x^2 \nonumber \right.\\
     &\left. \qquad -\frac{1}{4} \partial_{xx}w(x) \Delta x^3-\frac{1}{12} \partial_{xxx}w(x) \Delta x^4 \right\rangle + \mathcal{O}(\Delta x^5)~.
\end{align}
To find the $\mathcal{O}(\Delta t^2)$ contribution in the above we need to use Eq. \eqref{app_eqn:secondodlagevin} to find:
\begin{align*}
    \expected{\Delta x|x,t} &\sim \Phi(x,t) \Delta t + \frac{1}{2} \Phi(x,t) \partial_{x}\Phi(x,t) \Delta t^2 \\
    & \qquad +\ot{3} ~,\\
    \expected{\Delta x^2|x,t} & \sim \sigma^2 \Delta t +  \Phi(x,t)^2\Delta t^2 + \sigma^2 \partial_{x}\Phi(x,t) \Delta t^2 \\
    & \qquad +\ot{3} ~,\\
    \expected{\Delta x^3|x,t} & \sim 3\sigma^2 \Phi(x,t)  \Delta t^2+\ot{3} ~,~
    \text{and}\\ 
    \expected{\Delta x^4|x,t} & \sim \sigma^4 \Delta t^2 +\ot{3}
    ~.
\end{align*}
Substitution of these terms into the expansion yields the following coefficient for the $\ot{2}$ term: 
\begin{align*}
 C_{2}^{u}(x,t)
    & = \frac{1}{2}\partial_{x}w(x)[\Phi(x,t)^2+\sigma^2 \partial_{x}\Phi(x,t)]\nonumber \\
    & \qquad +\frac{1}{4}\partial_{xx}w(x) [3\sigma^2 \Phi(x,t)]+\frac{1}{12}\partial_{xxx}w(x) \sigma^4~.
\end{align*}
\section{Models and simulation details}

The following discusses a specific class of overdamped Langevin systems---linear models. Their Langevin equations have the form of:
\begin{align}\label{app_eqn:linearlangevin}
    \boldsymbol{z}= A \cdot \boldsymbol{z} dt + \sqrt{2 \boldsymbol{D}}\cdot d\boldsymbol{W}_{t}
    ~,
\end{align}
where $\boldsymbol{z}\in \mathbb{R}^{n}$ is the state vector, $A\in \mathbb{R}^{n\times n}$ is a matrix that defines the deterministic part of the Langevin dynamics, $\mathbf{D}$, a positive diagonal matrix, is the diffusion constant matrix, and $d\boldsymbol{W}_{t}\in \mathbb{R}^{n}$ is an $n-$dimensional infinitesimal Wiener process.

The bead-spring model, in both underdamped and overdamped regimes, falls within the class of linear models. One of their notable properties is that if the initial probability distribution is Gaussian, it remains Gaussian throughout the evolution.

To see this, suppose the state vector at time $t$, $\boldsymbol{z}(t)$ obeys Gaussian distribution, i.e., $\boldsymbol{z}(t)\sim \mathcal{N}(\boldsymbol{\mu}(t),\Sigma(t))$ where $\boldsymbol{\mu}(t)$ and $\Sigma(t)$ are the mean vector and the covariance matrix at time $t$, respectively. The state vector $\boldsymbol{z}$ at time $t+dt$ is $\boldsymbol{z}(t+dt)=(1+A dt)\cdot\boldsymbol{z} + \sqrt{2 \boldsymbol{D}}\cdot d\boldsymbol{W}_{t} (t)$. We see that $z(t+dt)$ also obeys Gaussian distribution using the following facts: 
\begin{enumerate}
    \item If $X$ is a multivariate normal random variable and $A$ is a constant matrix, then $A\cdot X$ is a multivariate normal random variable; and
    \item If $X_{1}$ and $X_{2}$ are independent multivariate normal random variables, their sum $X_{1}+X_{2}$ is also a multivariate normal random variable.
\end{enumerate}
Thus, to solve any linear model initialized from a Gaussian, we only need to solve how $\boldsymbol{\mu}$ and $\Sigma$ evolve with time $t$. To solve $\boldsymbol{\mu}$, we take average of Eq. \eqref{app_eqn:linearlangevin}:
\begin{align}
    d \boldsymbol{\mu} = A \cdot \boldsymbol{\mu}dt
    ~,
\end{align}
which is simply a set of linear ordinary differential equations. The solution can be written as:
\begin{align}
    \boldsymbol{\mu}(t) = \exp{(A t)}\cdot \boldsymbol{\mu}(0)
    ~.
\end{align}
For the covariance matrix $\Sigma = \expected{(\boldsymbol{z}-\boldsymbol{\mu})(\boldsymbol{z}^{\top}-\boldsymbol{\mu}^{\top})}$, the infinitesimal change is:
\begin{align*}
    & \Sigma(t+dt) -\Sigma(t) \\
    &= \expected{(\boldsymbol{z}-\boldsymbol{\mu}+d\boldsymbol{z}-d\boldsymbol{\mu})(\boldsymbol{z}^{\top}-\boldsymbol{\mu}^{\top}+d\boldsymbol{z}^{\top}-d\boldsymbol{\mu}^{\top})} \\
    &\qquad -\expected{(\boldsymbol{z}-\boldsymbol{\mu})(\boldsymbol{z}^{\top}-\boldsymbol{\mu}^{\top})} \\
    &=\expected{(\boldsymbol{z-\boldsymbol{\mu}})(d\boldsymbol{z}^{\top}-d\boldsymbol{\mu}^{\top})} +\expected{(d\boldsymbol{z}-d\boldsymbol{\mu})(\boldsymbol{z}^{\top}-\boldsymbol{\mu}^{\top})}
     \\
    &\qquad +\expected{d\boldsymbol{z}d\boldsymbol{z^{\top}}} \\
    &=\expected{(\boldsymbol{z-\boldsymbol{\mu}})(\boldsymbol{z}^{\top}-\boldsymbol{\mu}^{\top})\cdot A^{\top}}dt \\
    & \qquad +\expected{A\cdot(\boldsymbol{z}-\boldsymbol{\mu})(\boldsymbol{z}^{\top}-\boldsymbol{\mu}^{\top})}dt + 2 \boldsymbol{D}dt
    ~.
\end{align*}
This yields the time derivative of the covariance matrix:
\begin{align}
    \frac{d}{dt} \Sigma=\Sigma\cdot A^{\top}+A\cdot\Sigma+2\boldsymbol{D}
    ~. 
\end{align}
The stable distribution can be found by solving equations:
\begin{align}\label{app_eqn:stable_distribution}
    \Sigma\cdot A^{\top}+A\cdot\Sigma+2\boldsymbol{D}=0
    ~.
\end{align}

In the simulation, we generate timesteps using Euler–Maruyama method according to Eq. \eqref{app_eqn:linearlangevin}. The $dt$ is set to be $0.01$. We set three different initial distributions $\boldsymbol{x}(t=0)\sim \mathcal{N}(\boldsymbol{0}, \Sigma_{1}) $, $\boldsymbol{x}(t=0)\sim \mathcal{N}(\boldsymbol{0}, \Sigma_{2}) $ and $\boldsymbol{x}(t=0)\sim \mathcal{N}(\boldsymbol{0}, \Sigma_{\text{NESS}})$ where:
\begin{align}
\Sigma_{1}&=\mathrm{diag}(1,2,3,4,5),\\
\Sigma_{2}&=
\begin{pmatrix}
5  & 2  & 1  & -1 & 1  \\
2  & 8  & -1 & 1  & -2 \\
1  & -1 & 9  & -2 & 1  \\
-1 & 1  & -2 & 7  & -1 \\
1  & -2 & 1  & -1 & 8  
\end{pmatrix}, ~\text{and}\\
\Sigma_{\text{NESS}}&=
\begin{pmatrix}
\frac{1823}{1980} & \frac{833}{990} & \frac{1361}{1980} & \frac{959}{1980} & \frac{1}{4} \\
\frac{833}{990} & \frac{58}{33} & \frac{2819}{1980} & 1 & \frac{1021}{1980} \\
\frac{1361}{1980} & \frac{2819}{1980} & \frac{9}{4} & \frac{3121}{1980} & \frac{1609}{1980} \\
\frac{959}{1980} & 1 & \frac{3121}{1980} & \frac{74}{33} & \frac{1147}{990} \\
\frac{1}{4} & \frac{1021}{1980} & \frac{1609}{1980} & \frac{1147}{990} & \frac{3127}{1980} 
\end{pmatrix}
~.
\end{align}
$\Sigma_{\text{NESS}}$ is the NESS covariant matrix at $T_{h}/T_{c}=2$.
\section{Different initial distributions}\label{app:results_diffdis}
\begin{figure*}
   \centering
    \begin{subfigure}{1.0\columnwidth}
        \centering
        \caption{}
        \includegraphics[width=\linewidth]{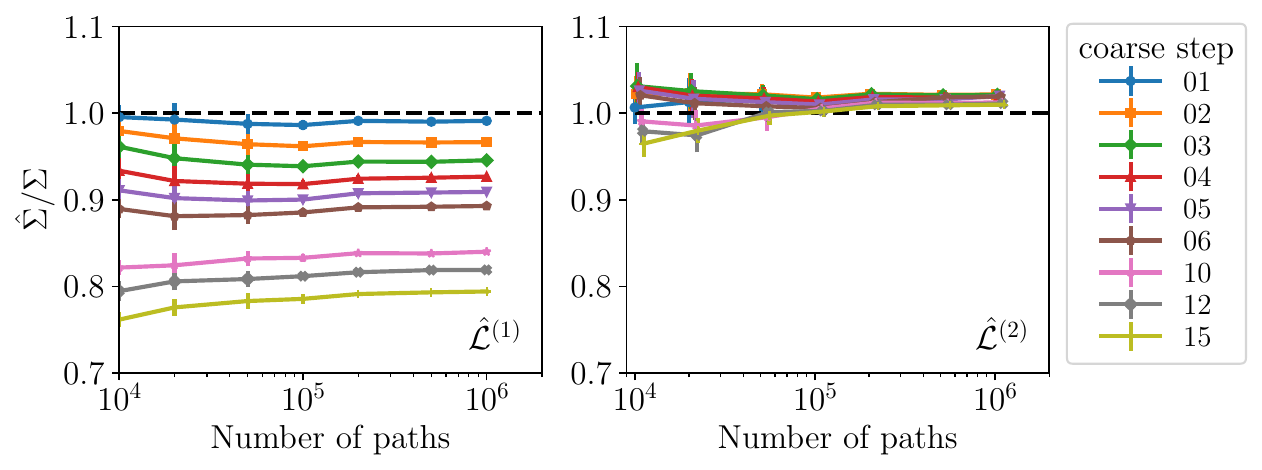}
        \label{fig:ent_temp_dis3}
    \end{subfigure}
    \hfill
    \begin{subfigure}{1.0\columnwidth}
        \centering
        \caption{}\includegraphics[width=\linewidth]{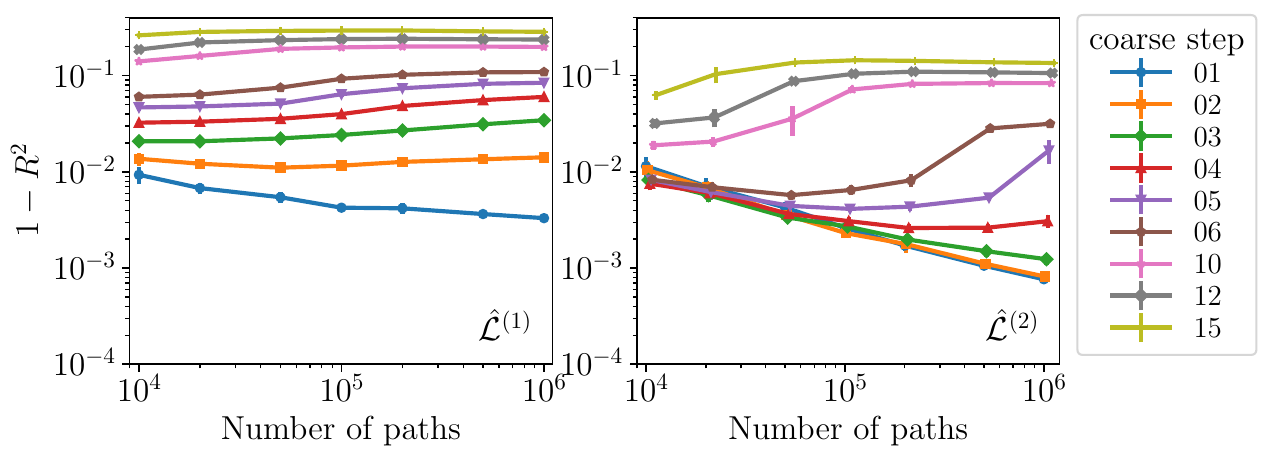}
        \label{fig:r2_N_dis3}
    \end{subfigure}
\caption{Mean entropy production and $R^2$ score between theoretical and estimated stochastic entropy production with initial distribution $\Sigma_{2}$ at $T_{h}/T_{c}=2$: (a) Relative entropy production as a
function of the number of paths using different coarse steps. The left (right) plot is the relative entropy with the first- (second-)order loss functions. (b) $1-R^2$ score between trajectory’s theoretical and estimated stochastic entropy productions as functions of the number of paths. The left (right) plot is the result with the first- (second-)order loss functions.
}
\label{fig:results_dis3}
\end{figure*}

\begin{figure*}
   \centering
    \begin{subfigure}{1.0\columnwidth}
        \centering
        \caption{}
        \includegraphics[width=\linewidth]{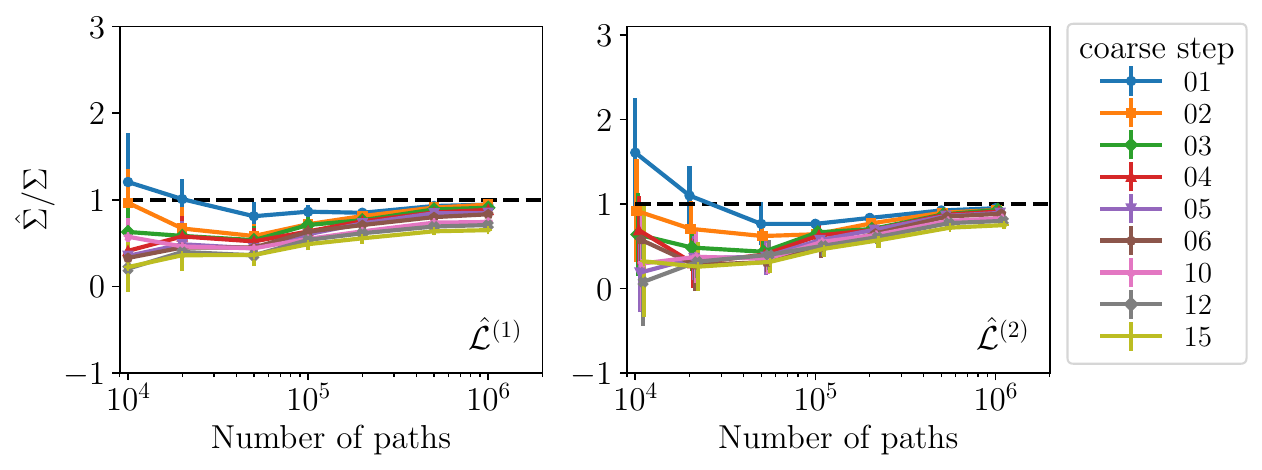}
        \label{fig:ent_N_ness}
    \end{subfigure}
    \hfill
    \begin{subfigure}{1.0\columnwidth}
        \centering
        \caption{}\includegraphics[width=\linewidth]{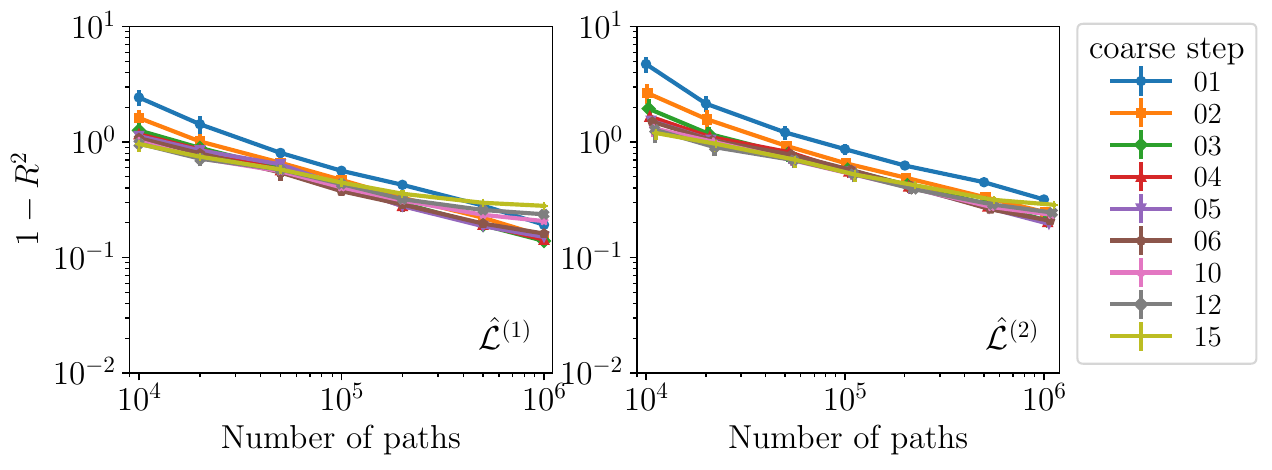}
        \label{fig:r2_N_ness}
    \end{subfigure}

\caption{Mean entropy production and $R^2$ score between the theoretical and estimated stochastic entropy production with $\Sigma_{\text{NESS}}$ at $T_{h}/T_{c}=2$: (a) Relative entropy production as a
function of the number of paths using different coarse steps. The left (right) plot gives the relative entropy results with the first- (second-)order loss functions. (b) $1-R^2$ score between trajectory’s theoretical and estimated stochastic entropy productions as functions of the number of paths. The left (right) plot gives the results with the first- (second-)order loss functions.
}
\label{fig:results_ness}
\end{figure*}

The main text presented the stochastic entropy production results for the five-bead spring system where the initial distribution is Gaussian with a diagonal covariance matrix. This section illustrates more results with different initial distributions. 

We first consider an initial Gaussian distribution that has nonzero off-diagonal terms in the covariance matrix. The results are showen in Figs. \ref{fig:results_dis3} and \ref{fig:results_ness}. With the initial distribution $\Sigma_{2}$, we see nonmonotonic behavior in the curves with coarse-step exceeding $4$ with the second-order loss function. $R^2$ in this case is less tolerant of the coarse step. For the NESS, unlike the estimation with non-NESS data, we find that NESS data with different coarse steps leads to similar average entropy-production estimation results. And, the second -rder estimation results are close to the first order. In term of $R^2$, the coarse-grained data produces even better results.

\section{Underdamped Systems}\label{app:underdamped}

After focusing on overdamped systems, we now turn to underdamped.

\subsection{Dynamics}

Consider a one-dimensional underdamped dynamic governed by:
\begin{align}\label{eqn:udeqns}
    dx &= vdt \nonumber \\
    mdv &= F(x,v,t) dt + \sigma dW_{t} \nonumber \\
    &=-\gamma vdt + f(x,t)dt + \sigma dW_{t}~, 
\end{align}
where $m$ and $\gamma$ are the particle mass and damping coefficient, $F(x,v,t)$ is the total force, including the linear damping force, and $f(x,t)$ is an external force independent of velocity $v$. The variance of the thermal noise $\sigma^2$ satisfies the fluctuation dissipation relation: $\frac{1}{2}\sigma^2 = \gamma T$, where $T$ is the temperature. The corresponding probability distribution $f(x,v,t)$ evolution is governed by the Klein-Kramers equation:
\begin{align}\label{app_eqn:kramereq}
   \partial_{t}f(x,v,t) + v\partial_{x}f(x,v,t) +\partial_{v}[f(x,v,t)u_{v}]=0
   ~,
\end{align}
where $u_{v}=\frac{1}{m}[F(x,v,t)-\frac{\gamma T}{m}\partial_{v}\log f(x,v,t)]$ is the probability velocity along the direction of $v$.

Before discussing estimation of underdamped dynamics, we define several stochastic integrals in the It\^{o} convention. These are important when addressing the approximation order we expect:
\begin{align} \label{app_eqn:stoch_identities}
    I_{w}^{(n)} &= \int_{t}^{t+n\Delta t }dW_{s} =W_{n\Delta t}\sim \mathcal{N}(0, n \Delta t)\\
    I_{0w}^{(n)} &= \int_{t}^{t+n\Delta t }ds\int_{t}^{s}dW_{s'}=\int_{0}^{n \Delta t}W_{s}ds \nonumber \\
    &\sim \mathcal{N}(0, \frac{1}{3}(n \Delta t)^3)\\
    I_{00}^{(n)} &=\int_{t}^{t+n\Delta t }ds\int_{t}^{s}ds' =\frac{1}{2}(n \Delta t)^2~.
\end{align} 
Also of importance are sevefral ensemble averages of the products of $I_{0w}^{(n)}$ given by: $\langle I^{(n)}_{0w} I^{(m)}_{0w}\rangle = \Delta t^{3} f_{nm} $, where $f_{11}=1/3$, $f_{12}=5/6$, and $f_{22}=8/3$.

Let's first learn the force $F(x,v,t)$ in the dynamics of Eq. \eqref{eqn:udeqns}---the analog to learning the drift $\Phi$ for overdamped systems. According to the method laid out in the main body, inspecting Eq. \eqref{eqn:udeqns} leads to expecting the loss function for learning the force: 
\begin{align}\label{eqn:cost_force_withxv}
    \mathcal{L}_{F}(x,v)=\expected{-w(x,v) F(x,v,t) dt+\frac{1}{2}w^2(x,v)dt}
    .
\end{align}
With the corresponding first-order discrete version as:
\begin{align}\label{app_eqn:loss_ud_discrete_with_xv}
    \mathcal{L}_{F}^{(1)}(x,v)=\expected{-w(x,v) \Delta v+\frac{1}{2}w^2(x,v)\Delta t}
    ~.
\end{align}

Then, we deploy a neural network $w(x,v)$ to learn the force at time $t$ with the loss function Eq. \eqref{app_eqn:loss_ud_discrete_with_xv}. This assumes access to the velocity via some kind of measurement. It is typical, however, to treat such a system as a partially observed---one for which we only have access to the time series of positions: $\{x(a\Delta t)\}_{a=0}^{N}$. In such a case, the velocity must be estimated. Our perspective is especially well suited to estimation of a partially-observed system---since the goal is to find alternative loss functions that agree upon averaging with loss functions that are experimentally inaccessible. In this case, we will assume the simplest estimation for the velocity:
\begin{align}\label{app_eqn:v_estimation}
    \widehat{v}(t) = \frac{x(t+\Delta t) - x(t)}{\Delta t}
    ~.
\end{align}
If we replace $v$ in Eq. \eqref{app_eqn:loss_ud_discrete_with_xv} with its estimation Eq. \eqref{app_eqn:v_estimation}, can we still learn the force $F(x,v,t)$ without changing anything else? The answer is no. However, appropriately modifying the loss function, we can learn the force. Simply replacing $v$ with $\widehat{v}$, the new loss function is: 
\begin{align}\label{app_eqn:loss_function_w/o_modi}
    \mathcal{L}_{F}(x,\hat{v})=\expected{-w(x,\hat{v})\Delta \hat{v}+\frac{1}{2}w^2(x,\hat{v})\Delta t}
    ~,
\end{align}
where $\Delta \widehat{v}=\widehat{v}(t+\Delta t)-\widehat{v}(t)$ and we remove the superscript ${(1)}$ since this is no longer the correct loss function.

We first compute the difference between $v$ and $\widehat{v}$ and then compute the average of the loss function Eq. \eqref{app_eqn:loss_function_w/o_modi} up to $\mathcal{O}(\Delta t)$. As in the overdamped case, we assume the data comes from a $\mathcal{O}(\Delta t^{3/2})$ integrator for the velocity:
\begin{align*}
&x(t+\Delta t) - x(t) = \int_t^{t+\Delta t} v(s) ds \\
&= \int_t^{t+\Delta t} ds\left( v(t) + \int_{t}^{s}F(x,v) ds' + \sigma W_{s} + \mathcal{O}(\Delta t^{3/2})\right)\\
&= v(t) \Delta t + \sigma I_{0w}^{(1)} + \int_{t}^{t+\Delta t} ds \int_{t}^{s}F(x,v) ds' + \mathcal{O}(\Delta t^{5/2})
~.
\end{align*}
The leading order of the force integral includes a factor of $I_{00}^{(1)}$, which is $\mathcal{O}(\Delta t^2)$, thus the difference between $\hat{v}$ and $v$ is given by:
\begin{align*}
    \widehat{v} = v+ \frac{1}{\Delta t}\sigma I_{0w}^{(1)} + F(x,v)\Delta t +\mathcal{O}(\Delta t^{3/2})~.
\end{align*}
We can now expand each term in the loss function. First, we address:
\begin{align}\label{app_eqn:vhat}
    \Delta \widehat{v} &=\widehat{v}(t+\Delta t)-\hat{v}(t) \nonumber \\
    & =  \frac{x(t+2 \Delta t)-2 x(t+\Delta t)+x(t)}{\Delta t} \nonumber \\
    &=  \frac{1}{\Delta t}(\sigma (I_{0w}^{(2)}-2I_{0w}^{(1)})) + F(x,v)\Delta t + \mathcal{O}(\Delta t^{3/2})~.
\end{align}
Similarly:
\begin{align*}
    &w(x,\widehat{v}) = w(x,v) + \partial_v w(x, v) \cdot (\widehat{v}-v) +\cdots\\
    &= w(x,v) + \partial_v w(x,v) \left[ \frac{1}{\Delta t}\sigma I_{0w}^{(1)} + F(x,v)\Delta t \right] \nonumber\\
    & \qquad + \frac{1}{2}\partial_{vv}w (\frac{1}{\Delta t} \sigma I_{0w}^{(1)})^2 +\ot{3}
    ~.
\end{align*}
Now, let's return to the loss function. The second term in Eq. \eqref{app_eqn:loss_function_w/o_modi} is already $\mathcal{O}(\Delta t)$ to leading order, so $\widehat{v}$ can be safely exchanged with $v$ in this term since the correction would be more than first order in $\Delta t$ and we are seeking a first-order loss function. The first term, however, has significant cross terms to contend with:
\begin{align*}
    w(x,\widehat{v})\Delta \widehat{v} &= w(x,v)\left[\frac{1}{\Delta t}(\sigma (I_{0w}^{(2)}-2I_{0w}^{(1)})) + F(x,v)\Delta t \right] \\ 
    & \quad + \partial_v w(x,v) \left[ \frac{1}{\Delta t^2 }\sigma^2 I_{0w}^{(1)} (I_{0w}^{(2)}-2I_{0w}^{(1)})\right]
    ~.
\end{align*}
The terms proportional to $I_{0w}^{(n)}$ in the first line vanish upon averaging. In the second line, we are terms proportional to $I_{0w}^{(n)}I_{0w}^{(m)}$. We eliminate these with the identities following Eq. \eqref{app_eqn:stoch_identities}. So, up to $\mathcal{O}(\Delta t)$, the loss function is:
\begin{align*}
    \mathcal{L}_{F}(x,\hat{v}) =&\expected{-w(x,v)F(x,v)\Delta t-\frac{1}{6}\sigma^2\partial_{v}w(x,v)\Delta t\\ 
    &+\frac{1}{2}w^2(x,v)\Delta t}~.
\end{align*}
This is not the correct loss function for learning force, so we need to modify the loss function to cancel out the appearance of the cross term:
\begin{align}\label{eqn:loss_force}
    \mathcal{L}_{F}^{(1)}(x,\widehat{v})&= \mathcal{L}_{F}(x,\widehat{v})+\expected{\frac{1}{6}\sigma^2\partial_{v}w(x,\widehat{v})\Delta t} \\
    &=\expected{-w(x,v)F(x,v)\Delta t+\frac{1}{2}w^2(x,v)\Delta t}
    ~. \nonumber
\end{align}
At first order $F(x,v)\Delta t \sim \Delta v$, so this loss function does agree with that using observed $v$. (Simply adding this extra term works out because any correction from replacing $\Delta t \partial_v w(x,\widehat{v})$ with $\Delta t \partial_v w(x,v)$ is higher order than $\Delta t$.) We can use this loss function to learn the force from the dynamics.

However, as we see in Eq. \eqref{eqn:loss_force}, there is a derivative term in the loss function. This sometimes makes the neural network hard to train. Again, by carefully tracking the $\Delta t$ order of our approximations, we can use another term to replace this derivative term. This is inspired by following relation, which is true for infinitesimal $dt$:
\begin{align}
   \expected{ dw \cdot \frac{dv}{dt} } = \expected{ \partial_v w\ dv\cdot \frac{dv}{dt}} =\expected{\sigma^2 \partial_v w}
   ~.
\end{align}

First, though, we must consider its finite expansion version: 
\begin{align*}
\Delta w &= w(x(t+\Delta t),\hat{v}(t+\Delta t),t+\Delta t)-w(x(t),\hat{v}(t),t)\\
    &=\partial_{t}w \Delta t + \partial_{x}w (v \Delta t + \sigma I_{0w}^{(1)}) \nonumber \\
    &+\partial_{v}w \cdot \frac{\sigma}{\Delta t}\left( I_{0w}^{(2)} - 2I_{0w}^{(1)}\right)+ \dots~.
\end{align*}

Now, let's consider the estimated version of the term above $\expected{\Delta w \frac{\Delta \widehat{v}}{\Delta t} }$. We know that $\Delta \widehat{v}/\Delta t$ has a leading order of $\Delta t^{-1/2}$. $\Delta w$ has a leading order of $\Delta t^{1/2}$, and its only $\mathcal{O}(\Delta t^{1/2})$ contribution is $\partial_{v}w [\frac{1}{\Delta t} \sigma(I_{0w}^{(2)}-2I_{0w}^{(1)})]$. Note that we only need $\mathcal{O}(\Delta t^0)$ since the term in question has a factor of $\Delta t$ attached. So, up to $\mathcal{O}(\Delta t^0)$ we have:
\begin{align*}
    \expected{\Delta w \frac{\Delta \hat{v}}{\Delta t}} &= \expected{\partial_{v}w \frac{\sigma^2}{\Delta t^3}[(I_{0w}^{(2)}-2I_{0w}^{(1)})][ (I_{0w}^{(2)}-2I_{0w}^{(1)})]} \nonumber \\
    & \qquad +\mathcal{O}(\Delta t^{1/2})\\
    &= \expected{\frac{2}{3} \sigma^2 \partial_{v}w}+\mathcal{O}(\Delta t^{1/2})
    ~,
\end{align*}
where in the second equality, we again use the ensemble average identities listed previously. Therefore, we can replace $\expected{\sigma^2 \partial_{v}w}$ in the loss function with $\frac{3}{2}\expected{\Delta w \frac{\Delta \hat{v}}{\Delta t}}$ without changing the order of our inference with respect to $\Delta t$. The loss function without the $w$ derivative is:
\begin{align}\label{app_eqn:lossfunction_force_wo_deri}
    &\mathcal{L}_{F}^{(1)}(x,\hat{v})\nonumber\\
    &=\expected{-w(x,\hat{v})\Delta \hat{v}+\frac{1}{2}w^2(x,\hat{v})\Delta t+\frac{1}{4} \Delta w(x,\hat{v}) \Delta \hat{v}}~.
\end{align}
As an additional bonus, the diffusion constant disappears in this loss function. There is no need to infer the diffusion constant before learning the force.

\subsection{Entropy production}

The average entropy production for an underdamped system is related to a local function---the \emph{irreversible velocity} \cite{spinney2012entropy}:
\begin{align*}
u_{irr}=-\gamma v/m-\gamma T/m^2\partial_{v}\log f
~.
\end{align*}
The mean entropy production rate is $m^2/ (\gamma T) \expected{u_{irr}^2}$. Thus, we must estimate the \jf for $u_{irr}$ rather then the full $u_v$ appearing in the Klein-Kramers equation. The derivation of this current appears in previous works \cite{lee2023multidimensional, lyu2024learning}, so we skip it here. The result is that to estimate the entropy production rate at time $t$ in underdamped dynamics, we  use the following loss function:
\begin{align*}
    \mathcal{L}_{u_{irr}}(x,v)= \expected{\frac{1}{2} w(x,v)^2dt - \frac{\gamma}{m} w(x,v) vdt + \frac{1}{2m} dw \cdot{dv}}
    ~.
\end{align*}
The minimum of this loss function leads the $u_{irr}$, i.e.: 
\begin{align}
    \argmin_{w} \mathcal{L}_{u_{irr}} = u_{irr}(x,v,t)
    ~.
\end{align}
Given access to the full phase space $(x,v)$, the corresponding discretized loss function can be used to estimate the entropy production rate:
\begin{align}
    \hatl_{u_{irr}}^{(1)}(x,v)= \expected{\frac{1}{2} w(x,v)^2\Delta t - \frac{\gamma}{m} w(x,v) v\Delta t + \frac{1}{2m} \Delta w \Delta v}
    ~.
\end{align}
However, when the velocity is not directly accessible, we need to consider the average of the following discretized version with $\widehat{v}$ and compute the modifications:
\begin{align*}
    \expected{\frac{1}{2} w(x,\hat{v})^2{\Delta t} - \frac{\gamma}{m} w(x,\hat{v}) \hat{v}{\Delta t} + \frac{1}{2m} \Delta w(x,\hat{v}) \Delta\hat{v}}
    ~.
\end{align*}
We expand up to $\mathcal{O}(\Delta t^{1})$. We can safely replace $\widehat{v}$ and $v$ in terms $\tfrac{1}{2}w(x,\hat{v})^2\Delta t$ and $\frac{\gamma}{m}w(x,\widehat{v})\hat{v}\Delta t$ as both are $\mathcal{O}(\Delta t^1)$ at leading order. For the term $\Delta w \Delta\widehat{v}$, we reuse the relation above:
\begin{align}
    \expected{\Delta w \frac{\Delta \widehat{v}}{\Delta t}} \sim \expected{\tfrac{2}{3} \sigma^2 \partial_{v}w}
    ~.
\end{align}
So we have:
\begin{align*}
    &\expected{\frac{1}{2} w(x,\hat{v})^2{\Delta t} - \frac{\gamma}{m} w(x,\hat{v}) \hat{v}{\Delta t} + \frac{1}{2m} \Delta w(x,\hat{v}) \Delta\hat{v}} \\
    &= \expected{\frac{1}{2} w(x,v)^2 {\Delta t}- \frac{\gamma}{m} w(x,v) v{\Delta t} + \frac{1}{3m}\sigma^2 \partial_{v}w(x,v) {\Delta t}}~.
\end{align*}
Recalling that $\sigma^2\partial_v w =dw \frac{dv}{dt} $ to leading order upon averaging, we see that this loss function is easily augmented: To learn the irreversible current correctly, the loss function should be:
\begin{align*}
    &\hatl_{u_{irr}}^{(1)}(x,\hat{v}) \\
    &=\expected{\frac{1}{2} w(x,\hat{v})^2{\Delta t} - \frac{\gamma}{m} w(x,\hat{v}) \hat{v}{\Delta t} + \frac{3}{4m} \Delta w(x,\hat{v}) \Delta\hat{v}}
    ~.
\end{align*}
This modified loss function facilitates learning the irreversible velocity in partially-observed underdamped dynamics.

\subsection{Diffusion constant}

Next, consider estimating the thermal noise variance $\sigma^2$ for underdamped dynamics. We use a network $w$ to learn $\sigma$. As in overdamped dynamics, the candidate loss function could be:
\begin{align}
\mathcal{L}_{\sigma}=\expected{\tfrac{1}{2}w^2 dt- w (dv)^2 }
~.
\end{align}
Or simply put, the variance of thermal noise is given by:
\begin{align}
    \sigma^2 dt = \expected{(dv)^2}
    ~.
\end{align}
With access to the velocity $v$, we use its discretized version:
\begin{align}
    {\sigma^2}^{(1)} = \expected{(\Delta v)^2/\Delta t} ~.
\end{align}
If we estimate the velocity $\widehat{v}$ using Eq. \eqref{app_eqn:vhat}, the average of our discretized version is:
\begin{align}
    \expected{( \Delta \hat{v})^2} = \frac{2}{3} \sigma^2 \Delta t + \mathcal{O}(\Delta t ^2)
    ~.
\end{align}
This leads to:
\begin{align}
    {\sigma^2}^{(1)} = \frac{3}{2}\expected{(\Delta \widehat{v})^2/\Delta t}
    ~,
\end{align}
which agrees with previous results \cite{pedersen2016connect, ferretti2020building}.

\subsection{Second-order loss functions for underdamped dynamics}

The main text's finite difference technique can be used to derive second-order loss functions for underdamped dynamics. We simply list the results directly, since there is no practical difference between these derivations and the one laid out in the main text:
\begin{align*}
\hatl_{F}^{(2)} & (x,\widehat{v}) \\
& = \expected{\tfrac{1}{2}w^2(x,\widehat{v})\Delta t-2w(x,\widehat{v})\Delta \widehat{v}+\tfrac{1}{2}w(x,\widehat{v})\Delta_{2} \widehat{v} \\
    &\qquad +\tfrac{1}{2}  \Delta w(x,\widehat{v}) \Delta \hat{v}-\tfrac{1}{8}\Delta_2 w(x,\widehat{v}) \Delta_2 \widehat{v}} \\
    &= \expected{\tfrac{1}{2}w^2(x,\widehat{v})\Delta t-2w(x,\widehat{v})\Delta \widehat{v}+\tfrac{1}{2}w(x,\widehat{v})\Delta_{2} \widehat{v} \\
    &\qquad +\tfrac{1}{4}  \Delta w(x,\widehat{v}) \Delta \widehat{v}} ~,
\end{align*}
\begin{align*}
    &\hatl_{u_{irr}}^{(2)}(x,\hat{v}) \\
    &=\expected{\tfrac{1}{2} w(x,\widehat{v})^2{\Delta t} - 2\frac{\gamma}{m} w(x,\widehat{v}) \Delta x + \tfrac{1}{2}\frac{\gamma}{m} w(x,\widehat{v}) \Delta_{2}x \\
    &\qquad + \tfrac{3}{2m} \Delta w(x,\widehat{v}) \Delta\widehat{v}-\frac{3}{8m} \Delta_{2} w(x,\widehat{v}) \Delta_{2}\widehat{v}} \\
    &=\expected{\tfrac{1}{2} w(x,\widehat{v})^2{\Delta t} - 2\tfrac{\gamma}{m} w(x,\widehat{v}) \Delta x + \tfrac{1}{2}\tfrac{\gamma}{m} w(x,\widehat{v}) \Delta_{2}x \\
    &\qquad + \tfrac{3}{4m} \Delta w(x,\widehat{v}) \Delta\widehat{v}}~,
\end{align*}
and:
\begin{align*}
    &{\sigma^2}^{(2)} = \tfrac{3}{2}\expected{2(\Delta \widehat{v})^2/\Delta t-\tfrac{1}{2}(\Delta_2 \widehat{v})^2/\Delta t}
    ~.
\end{align*}
We used the relation:
\begin{align}
    \Delta w(x,\widehat{v}) \Delta\widehat{v}\sim \frac{1}{2}\Delta_{2} w(x,\widehat{v}) \Delta_{2}\widehat{v}+\ot{2}
    ~.
\end{align}

\subsection{Trajectory-based entropy production}\label{app:ud_stochastic_ep}

This section derives the stochastic entropy production for an underdamped Langevin NESS for simplicity. The stochastic entropy production for each trajectory $\Gamma=\{(x_{t},p_{t})\}_{t=0}^{t=\tau}$ is straightforward to write down in Stratonovich notation:
\begin{align*}
    \sigma_{\Gamma} &= -\log f(x_{t},v_{t}) + \log f(x_{0},v_{0}) + Q/T \nonumber \\
    &=\int_{\Gamma} -\partial_{x} \log f \circ dx-\partial_{v} \log f \circ dv 
    \nonumber\\
    &\qquad + \frac{1}{T}(m\frac{dv}{dt}-\Phi)\circ dx \nonumber\\
    &=\int_{\Gamma}[-\partial_{x}\log f+ \frac{1}{T}(m\frac{dv}{dt}-\Phi)]\circ dx -\partial_{v} \log f \circ dv~\nonumber\\ 
    &= \int_{\Gamma}(-\partial_{x}\log f- \frac{1}{T}\Phi)v dt+ (\frac{1}{T}m v -\partial_{v} \log f )\circ dv~.   
\end{align*}

Unlike overdamped case, we must learn three functions from trajectories: $v\partial_{x} \log f$, $\partial_{v} \log f$, and $\Phi$. We already discussed how to infer the drift $\Phi$ and $\partial_{v}\log f$, so we only need to address the single remaining term. As before, we begin by assuming that we have full access to the phase space and then consider the partially-observed case.

To infer the $v\partial_{x}\log f$, we use the following current:
\begin{align*}
    \mathcal{L}_{vd_{x}\log f}=  \expected{\tfrac{1}{2}w^2(x,v)dt-w(x,v)v\partial_{x} \log fdt}~.
\end{align*}
Integration by parts leads to:
\begin{align*}
    \mathcal{L}_{vd_{x}\log f}=  \expected{\tfrac{1}{2}w^2(x,v)dt+v\partial_{x}w(x,v)dt}
    ~.
\end{align*}
The corresponding discrete version is:
\begin{align*}
    \hatl^{(1)}_{vd_{t}\log f}=\expected{\tfrac{1}{2}w^2(x,v)\Delta t + [w(x+\Delta x,v)-w(x,v)]}
    ~,
\end{align*}
where we used $w(x+\Delta x,v)-w(x,v) = \partial_{x}w(x,v) v \Delta t+\ot{2}$.

Thus, given full observational access to position and velocity, we can estimate the stochastic entropy production along a trajectory. Next, let us study the stochastic entropy production with access to position only.

To begin with, we write down the discretized version of stochastic entropy production for one step:
\begin{align}\label{app_eqn:dis_ud_stochastic_ep}
    \sigma_{\Gamma} & =[-\partial_{x}\log f(x, v)-\frac{1}{T}\Phi(x,v)]v\Delta t\nonumber \\
    &+\{\frac{1}{T}m\frac{v+(v+\Delta v)}{2}-\frac{1}{2}[\partial_{v}\log f(x+\Delta x, v+\Delta v)\nonumber\\
    &+\partial_{v}\log f (x,v)]\}\Delta v-\partial_{t}\log f(x,v)\Delta t
\end{align}

 
Now, we must write this in terms of the estimated velocity $\widehat{v}$. We find that, unlike the overdamped case, the stochastic entropy production of a trajectory cannot be computed from the trajectory and the local thermodynamic function only. This is due to the estimation of the velocity $\widehat{v}$. While the average entropy production can be accurately estimated without measuring the velocity directly, the trajectorywise entropy production cannot. There is an inherently stochastic term in the discretized stochastic entropy production expression ,if we must estimate the velocity.

To illustrate this point, we first look at a trivial current:
\begin{align}
    J= \int_{\Gamma}1 \circ dv
    ~.
\end{align}
The corresponding one-step discretized value $\Delta v$ and the estimated value $\Delta \widehat{v}$ are:
\begin{align*}
    \Delta v &= F(x,v)\Delta t+\sigma W_{\Delta t}+ \ot{3/2}  \\
    \Delta \hat{v} &=  \frac{1}{\Delta t}(\sigma (I_{0w}^{(2)}-2I_{0w}^{(1)})) + F(x,v)\Delta t + \ot{3/2}~,
\end{align*}
which gives:
\begin{align*}
    \Delta v = \Delta \hat{v} - \frac{1}{\Delta t}(\sigma (I_{0w}^{(2)}-2I_{0w}^{(1)}))+\sigma W_{\Delta t}~ +\ot{3/2}.
\end{align*}
In a long trajectory with $N$ steps, the relation between the actual current value and the estimated value is:
\begin{align}\label{app_eqn:trivial_current_stoch}
    J_{\Gamma} = \hat{J}_{\Gamma} + N [- \frac{1}{\Delta t}(\sigma (I_{0w}^{(2)}-2I_{0w}^{(1)}))+\sigma W_{\Delta t}]+\ot{1/2}
    ~.
\end{align}

From this example, with only the position data, there is a stochastic term in the difference between the estimated and actual current value. These stochastic terms have a vanishing ensemble average, meaning that we cannot learn them. We will only be able to learn $\widehat{J}_\Gamma$, rather than the true current $J_\Gamma$. Thus, with a partially observed current, we can only estimate the true current up to some variance.

With this intuition, we return to the entropy production. First, consider a general discrete one-step Stratonovich current of the form $w(x,v)\circ\Delta v$. For the fully and partially observed cases, we have the current values:
\begin{align*}
    &w(x,v) \circ \Delta v \\
    &= w(x,v)F(x,v) \Delta t + w\sigma W_{\Delta t} +\partial_{v}w(x,v) \tfrac{1}{2}\sigma^2W_{\Delta t}^2 
\end{align*}
and:
\begin{align*}
    &w(x,\widehat{v}) \circ\Delta \widehat{v} \\
    & = \tfrac{1}{2}[w(x,\widehat{v})+w(x+\Delta x,\widehat{v}+\Delta\widehat{v})]\Delta \widehat{v} \\
    &=w(x,v)F(x,v) \Delta t +w(x,v)\tfrac{1}{\Delta t}\sigma(I_{0w}^{(2)}-2I_{0w}^{(1)}) \\
    &\quad +\partial_{v}w\tfrac{1}{\Delta t^2}\sigma^2 I_{0w}^{(2)}(I_{0w}^{(2)}-2I_{0w}^{(1)})+\ot{3/2}
    ~.
\end{align*}
The relation between the fully and partially observed currents is:
\begin{align*}
& w(x,\widehat{v})\circ \Delta \widehat{v} \\
    & \quad =w(x,v) \circ \Delta v  -\sigma w(x,v)[ W_{\Delta t}-\frac{1}{\Delta t}(I_{0w}^{(2)}-2I_{0w}^{(1)})] \\
    & \quad -\sigma^2\partial_{v}w(x,v)[\frac{1}{2} W_{\Delta t}^2-\frac{1}{\Delta t^2}\ I_{0w}^{(2)}(I_{0w}^{(2)}-2I_{0w}^{(1)})] \\
    & \quad +\ot{3/2}~.
\end{align*}
With this relation, the entropy production in Eq. \eqref{app_eqn:dis_ud_stochastic_ep} can be written in terms of $\widehat{v}$:
\begin{align*}
\sigma_{\Gamma} & =[-\partial_{x}\log f(x, \widehat{v})-\frac{1}{T}\Phi(x,\widehat{v})]\widehat{v}\Delta t \\
    &+\{\frac{1}{T}m\frac{\widehat{v}+(\widehat{v}+\Delta \widehat{v})}{2}-\tfrac{1}{2}[\partial_{v}\log f(x+\Delta x, v+\Delta \hat{v}) \\
    &+\partial_{v}\log f (x,\hat{v})]\}\Delta \widehat{v} \\
    &+[\sigma \tfrac{m}{2T}\widehat{v}-\partial_{v}\log f(x,\widehat{v})][W_{\Delta t}-\tfrac{1}{\Delta t}(I_{0w}^{(2)}-2I_{0w}^{(1)})]  \\
    &+\sigma^2\partial_{vv}\log f(x,\widehat{v})[\tfrac{1}{2} W_{\Delta t}^2-\tfrac{1}{\Delta t^2}\ I_{0w}^{(2)}(I_{0w}^{(2)}-2I_{0w}^{(1)})] \\
    &+\ot{3/2}~.
\end{align*}
Echoing Eq. \eqref{app_eqn:trivial_current_stoch}, we think of the above as:
\begin{align*}
\sigma_{\Gamma} & = \hat{\sigma}_{\Gamma} \\
    &+[\sigma \frac{m}{2T}\hat{v}-\partial_{v}\log f(x,\hat{v})][W_{\Delta t}-\frac{1}{\Delta t}(I_{0w}^{(2)}-2I_{0w}^{(1)})] \\
    &+\sigma^2\partial_{vv}\log f(x,\hat{v})[\frac{1}{2} W_{\Delta t}^2-\frac{1}{\Delta t^2}\ I_{0w}^{(2)}(I_{0w}^{(2)}-2I_{0w}^{(1)})] \\
    &+\ot{3/2}
    ~.
\end{align*}

Again, we find that $\sigma_\Gamma$ can only be estimated to within a stochastic contribution that vanishes upon averaging. And so, it cannot be learned. For short times, we cannot expect this stochastic term to converge to its mean (and, thus, vanish) along a particular trajectory. Moreover, we will not be able to accurately estimate the entropy production. That said, for long trajectories, the stochastic terms converge to their corresponding means. Rather than a weakness of our method, we believe this is an illuminating point---one that's indicative of the generality of our approach to finding loss functions. Even in the case where a target function cannot be determined from the available observations, we are able to arrive at a loss function that captures as much as possible and quantitatively represents the uncertainty inherent in the partially-observed system by giving expressions from which we can estimate the uncertainty imposed by partial observations. 

\section{Markovian Jump Process}\label{app:Markovian}

Our $L^{2}-$based method also works for discrete Markovian jump processes. We consider the dynamics of an $n$-state Markovian jump process governed by the master equation:
\begin{align*}
    \frac{\partial{p}_{x}}{\partial t} =\sum_{y}R_{xy}p_{y} ~,
    \label{app_eqn: master equation}
\end{align*}
where $p_{x}(t)$ is the probability distribution on $n$ states and $R_{xy}$ is the transition-rate matrix with diagonal elements $R_{xx} = -\sum_{y, y\neq x} R_{yx}$. A stochastic trajectory $X_\tau$ can be denoted by a sequence of $N$ jump events:
\begin{align}
    \Gamma = [(x_0, t_0), (x_1, t_1) \dots,(x_N, t_N)  ],
\end{align}
where the system initiates at the state $x_0$ and undergoes a series of $N$ transitions from state $x_{n-1}$ to $x_{n}$ at times $t_n$ with $n=1,2,\cdots, N$. The entropy production of this Markovian jump process is related to the \emph{generalized thermodynamic force}:
\begin{align*}
f_{xy}=\log\frac{R_{xy}p_{y}}{R_{yx}p_{x}}
~.
\end{align*}
The short-time entropy production is:
\begin{align}
    \sigma dt=\sum_{(x,y)}R_{xy}p_{y} f_{xy}dt
    ~.
\end{align}

To estimate the entropy production, we learn the function $f_{xy}$. According to our method, this means using a loss function:
\begin{align*}
\mathcal{L}_f = \E[\frac{1}{2} w^2_{xy} - f_{xy}w_{xy}]
~,    
\end{align*}
where the expectation is taken over a data set of observed jumps starting from state $x$ and ending in state $y$. Since the process is Markovian by definition, we do not keep track of where the state was before it landed in state $x$. As such, the data set can be reorganized into a set of $\mathcal{D} = \{n_{xy}\}$, where $ n_{xy}$ is the number of transitions from state $y$ to state $x$ during time duration from $t$ to $t+dt$. For this data set, rather than sum over observed jumps, we sum over pairs of states:
\begin{align*}
\mathcal{L}_f = \E[\sum_{x\neq y}\left(\tfrac{1}{2} w^2_{xy}  - f_{xy}w_{xy}\right)n_{xy}]
~.
\end{align*}

This motivates introducing a current with weight $w_{xy}$---the \emph{empirical current}:
\begin{align*}
j(w_{xy}) dt := \sum_{x\neq y}w_{xy}n_{xy}
~.
\end{align*}
The expectation value of this short time current with weight $w_{xy}$ is:
\begin{align}
    \expected{\sum_{x\neq y}w_{xy}n_{xy}}=\sum_{x\neq y} w_{xy}R_{xy}p_{y}dt
    ~.
\end{align}

From this perspective, we see that $\sigma dt = \E[j(f_{xy}) dt]$ and the weight $R_{xy}p_{y}$ in the expected value expression are analogous to the weight $u(x,t)$ appearing in the Stratonovich current in overdamped dynamics. However, note that it does not appear in the entropy production the way that $u$ did.

Using notation from the main development, the expected values of short empirical currents with weights $w^2$ and $wf$ (analogs of the \jw and \jf for Markov jump processes) are:
\begin{align*}
    \expected{j_{w}}dt & = \sum_{x\neq y}\frac{1}{2} w_{xy}^2R_{xy}p_{y}dt ~\text{and} \\
    \expected{j_{f}}dt & = \sum_{x\neq y} w_{xy}f_{xy}R_{xy}p_{y}dt
    ~.
\end{align*}
With some foresight, instead of learning generalized thermodynamic force directly, we choose to learn the function:
\begin{align*}
r_{xy}=e^{ f_{yx}}=\frac{R_{yx}p_{x}}{R_{xy}p_{y}}
~. 
\end{align*}
To do this, we expect the loss function to be:
\begin{align*}
    \mathcal{L}_{r_{xy}} & =\expected{j_{w}dt}-\expected{j_{r}dt} \\
    & =\sum_{x\neq y}(\frac{1}{2}w_{xy}^2-r_{xy}w_{xy})R_{xy}p_{y}dt
    ~.
\end{align*}
To estimate the correlation current from data, we note that:
\begin{align*}
    \expected{j_{r}dt}& =\sum_{x\neq y}w_{xy}\frac{R_{yx}p_{x}}{R_{xy}p_{y}}R_{xy}p_{y}dt \\
    & =\sum_{x\neq y}w_{xy}{R_{yx}p_{x}}dt \\
    & =\sum_{x\neq y}1/w_{xy}{R_{xy}p_{y}}dt
    ~,
\end{align*}
where, in the last equality, we impose the constraint $w_{xy}w_{yx}=1$ for any pair of $(x,y)$. Thus, the loss function can be estimated from:
\begin{align}
    \widehat{\mathcal{L}}_{r_{xy}}=\expected{\sum_{x\neq y}(\frac{1}{2}w_{xy}^2-\frac{1}{w_{xy}})n_{xy}}
    ~,
\end{align}
provided we impose the symmetry constraint $w_{xy}w_{yx}=1$ on the weight function. This loss function can be estimated from the data directly and the minimum leads to:
\begin{align*}
w_{xy} = r_{xy} = \frac{R_{yx}p_{x}}{R_{xy}p_{y}}
~.
\end{align*}
The stochastic entropy production for a trajectory is

\begin{align}
    \sigma_{\Gamma} = -\int_{\Gamma}\partial_{t}\log p_{x}dt+\sum_{t_{i}} \log \frac{R_{x_{i+1}x_{i}}p_{x_{i}}}{R_{x_{i}x_{i+1}}p_{x_{i+1}}}\Theta(t-t_{i}),
\end{align}
where the first contribution comes from dwelling and the second from jumps. To learn the temporal score function $\partial_{t} \log p_x$, simply consider a state function $w_{x}$ and loss function
\begin{align}
    &\mathcal{L}_{\partial_{t}\log p}=\E[\frac{1}{2}w_{x}^2 dt- w_x \partial_{t}\log p_x dt] \nonumber\\
    &= \expected{\frac{1}{2}w_{x}^2dt}-d\expected{w_{x}},
\end{align}
where we use $\expected{ w_x \partial_{t}\log p_xdt}=\sum_{x}w_{x}\partial_{t} p_xdt=d\expected{w_{x}} $. Here, $d\expected{w_{x}}$ is the change in expected $w_{x}$ from time $t$ to $t+dt$.
After these, we can learn the stochastic entropy production in Markovian jump processes.
\end{document}